\documentclass[draft]{agujournal2019}
\usepackage{url} 
\usepackage[inline]{trackchanges} 
\usepackage{soul}
\usepackage{epstopdf}
\usepackage{ragged2e} 
\usepackage[utf8]{inputenc}
\usepackage{algorithm}
\usepackage{algpseudocode} 
\usepackage{graphicx} 
\usepackage{subcaption} 
\usepackage{xr} 
\usepackage{natbib}
\usepackage{xcolor}

\draftfalse

\journalname{JGR: Machine Learning and Computation}

\begin{document}

%
%


\title{Earthquake Source Depth Determination using Single Station Waveforms and Deep Learning }

%
%




\authors{Wenda Li\affil{1}, Miao Zhang\affil{1}}

 \affiliation{1}{Department of Earth and Environmental
Sciences, Dalhousie University, Halifax, Nova Scotia, Canada}






\correspondingauthor{Miao Zhang}{Miao.Zhang@dal.ca}


 \begin{keypoints}
 \item We developed VGGDepth, a deep learning model for earthquake depth estimation using single-station three-component waveforms.
 \item	
 VGGDepth achieves depth error of less than 1km in both single-station and regionally generalized models, with smaller errors when averaging.
 \item	
VGGDepth proves little dependence on the nearest stations, offering big potential for application with single stations or sparse networks.
 \end{keypoints}

%
%

%
%

\begin{justify}
\begin{abstract}

In areas with limited station coverage, earthquake depth constraints are much less accurate than their latitude and longitude. Traditional travel-time-based location methods struggle to constrain depths due to imperfect station distribution and the strong trade-off between source depth and origin time. Identifying depth phases at regional distances is usually hindered by strong wave scattering, which is particularly challenging for low-magnitude events. Extracting effective depth features from single or sparse stations to enhance depth constraints is a pressing challenge. Deep learning algorithms, capable of extracting various features from seismic waveforms, including phase arrivals,  amplitudes, and frequency, offer promising constraints to earthquake depths. In this work, we propose a novel depth feature extraction network (named VGGDepth), which directly maps seismic waveforms to earthquake depth using single-station three-component waveforms. The network structure is adapted from VGG16 in computer vision. It is designed to take single-station three-component waveforms as inputs and produce depths as outputs. Two scenarios are considered in our model development: (1) training and testing solely on the same station, and (2) generalizing by training and testing on different seismic stations within a particular region. We demonstrate the efficacy using seismic data from the 2016-2017 Central Apennines, Italy earthquake sequence. Results demonstrate that earthquake depths can be estimated from single stations with uncertainties of hundreds of meters. These uncertainties are further reduced by averaging results from multiple stations. Our method shows strong potential for earthquake depth determination, particularly for events recorded by single or sparsely distributed stations, such as historically instrumented earthquakes.

\end{abstract}

\section*{Plain Language Summary}

Understanding earthquake depths is crucial for both scientific research and building safety. However, current methods are limited in their ability to fully extract earthquake depth information from seismic waveform data. To address this limitation, we developed a new deep learning based method called VGGDepth, which estimates earthquake depth directly from a single seismic station. Additionally, we proposed and trained a regionally generalized model, enabling the method to use waveform data from any available stations within the region. We demonstrate the effectiveness of our approach using seismic data from the 2016-2017 Central Apennines, Italy earthquake sequence.

%
%

%


%
%
%
%

\section{Introduction}

Among earthquake hypocentral parameters, depth is generally less accurately constrained than epicentral location, especially for events recorded by sparse or distant stations. Small earthquakes dominate in earthquake catalogs, but they are often detected and located by only a few stations. Additionally, many historical seismic events were recorded by sparse stations --- sometimes even by a single station.  As a result, determining hypocentral depth under sparse station conditions remains a significant challenge \cite[]{gomberg1990effect, ma2011combining, zhang2014new}. Nevertheless, earthquake depth plays a crucial role in characterizing earthquakes \cite[]{zhang2021source}, understanding subsurface structures \cite[]{chandrasekhar2012new}, and assessing seismic hazards \cite[]{sharma1991underground}. For example, shallow earthquakes often pose greater seismic hazards and can be significantly more destructive than deeper events, as highlighted by the 1993 Latur earthquake in India \cite[]{gupta1996fluids}, which had a magnitude of 6 and a depth of just 2 km, resulting in nearly 10,000 fatalities. Therefore, depth is a core parameter in Probabilistic Seismic Hazard Analysis (PSHA), essential for improving its accuracy and serving as a critical link between earthquake characterization and seismic risk \cite[]{richards2006applicability, petersen2008documentation, gupta2013source}. 

Travel-time-based methods have been the predominant approach for earthquake location due to their efficiency and accuracy (e.g., VELEST: \citeauthor{kissling1995program}, \citeyear{kissling1995program}; HypoInverse: \citeauthor{klein2002user}, \citeyear{klein2002user}; NonLinLoc: \citeauthor{lomax2014earthquake}, \citeyear{lomax2014earthquake}); however, earthquake depths are often poorly constrained using these methods, particularly under sparse station conditions. Theoretically, at least one S-phase arrival must be recorded by a station located at a distance of $<\ \sim$1.4 times the event depth to reliably constrain the depth \cite[]{gomberg1990effect}. Moreover, travel-time-based location methods suffer from a strong trade-off between source depth and origin time, especially when station coverage is sparse or nearby stations are lacking. High-precision relative location techniques can refine the relative locations between events (e.g., HypoDD: \citeauthor{waldhauser2000double},  \citeyear{waldhauser2000double}; COMPLOC: \citeauthor{lin2006comploc}, \citeyear{lin2006comploc}; GrowClust: \citeauthor{trugman2017growclust},  \citeyear{trugman2017growclust}), but they have limited capability to improve absolute locations. Although the machine-learning-based phase picking methods (e.g., PhaseNet: \citeauthor{zhu2019phasenet},     \citeyear{zhu2019phasenet}; EQTransformer: \citeauthor{mousavi2020earthquake},     \citeyear{mousavi2020earthquake}; OBSTransformer: \citeauthor{niksejel2024obstransformer},     \citeyear{niksejel2024obstransformer}) and associated location workflows  (e.g., LOC-FLOW: \citeauthor{zhang2022loc},     \citeyear{zhang2022loc}; QuakeFlow: \citeauthor{zhu2023quakeflow}, \citeyear{zhu2023quakeflow}) have significantly enhanced earthquake detection and location capabilities, depth estimations still fundamentally depend on the availability of nearby stations, much like traditional methods.

Identifying depth phases at regional distances provides effective and reliable constraints on earthquake depth, even under sparse station conditions. These phases may originate from surface-converted or Moho-reflected waves (e.g., sPL: \citeauthor{chong2010spl}, \citeyear{chong2010spl}; sPg, sPmP, sPn: \citeauthor{ma2011combining}, \citeyear{ma2011combining}). Such approaches can yield accurate depth estimates with uncertainties of less than 1 km but typically require complex waveform simulations that depend on velocity models and focal mechanisms. To bypass the need for waveform simulation or explicit depth-phase identification, \cite{zhang2014new} proposed a method that automatically extracts depth information by stacking multi-station autocorrelations of seismic body or coda waves. Similarly, \cite{yuan2020depth} developed the Depth-Scanning Algorithm, which identifies various depth phases through phase shifting and cross-correlation. At teleseismic distances, depth phases such as sP, pP, and pwP are relatively simpler and more easily identifiable for depth constraints (e.g., \citeauthor{engdahl1998global}, \citeyear{engdahl1998global}), and their identification has recently been advanced through deep learning techniques \cite[]{munchmeyer2024learning}. However, at regional distances and under sparse station conditions, depth-phase identification and waveform simulation remain challenging and vary regionally due to complex geological structures and strong wave scattering, which pose particular difficulties for low-magnitude events.

In addition to converted or reflected depth phases, other seismic phases and waveform characteristics are also sensitive to earthquake depth, such as surface phases and coda waves. Both surface waves and coda waves are more pronounced in shallow earthquakes than in deeper ones. For surface waves, two main approaches are commonly used. The first approach involves computing the amplitude ratios of surface waves to body waves (e.g., $Rg/Sg$: \citealp{langston1987depth}; \citealp{tibi2018depth}; \citealp{zhang2021source}). 
The second approach directly fits the spectral amplitudes of the observed and synthetic surface waves \cite[]{jia2017joint, he2023automatic, somiah2025reassessment}. Both approaches require significant waveform modeling by searching source depths. In contrast, the coda wave-based method qualitatively differentiates source depths by comparing the magnitude differences inferred from body waves and coda waves, which, as noted by \citet[]{holt2019portability} and \citet[]{voyles2020new}, is particularly useful for distinguishing between earthquakes and explosions. Recently, \citet{koper2024inferring} proposed to combine multiple features --- including magnitudes, source spectra, and phase ratios --- to improve depth discrimination. Although the above methods improve depth determination accuracy, their applications remain limited due to specific dependencies, such as waveform modeling, velocity models, and complex procedures like individual physical feature computation.

Deep learning algorithms, capable of extracting depth-related features from complete waveforms and leveraging travel time, amplitude, as well as frequency information, offer promising solutions for constraining earthquake depths. Mapping source parameters directly from waveforms eliminates many empirical criteria, significantly improving both the convenience and accuracy of depth estimation. \citet{lomax2019investigation} utilized the convolutional neural network to frame earthquake location as a classification problem using single-station waveforms. \citet{mousavi2019bayesian} and \citet{ noda2024deep} demonstrated that convolutional neural networks hold great potential for estimating epicentral distances from single-station waveforms. \citet{elsayed2023eqconvmixer} proposed the ConvMixer network, which enhanced the accuracy of seismic hypocenter location using single-station waveforms.
\citet{ristea2021complex} proposed a complex convolutional neural network architecture that directly estimates epicentral distance, depth, and magnitude from single-station waveforms, achieving depth uncertainties of a few kilometers. Additionally, deep learning methods incorporating multiple stations have been developed. For example, \citet{zhang2020locating} designed a fully convolutional network for earthquake location using multiple-station waveforms, and later improved it for real-time earthquake early warning \cite[]{zhang2021real, zhang2023generalized}. \citet{munchmeyer2021earthquake} proposed attention-based transformer networks for real-time estimation of earthquake magnitude and location. \citet{zhang2022spatiotemporal} proposed a Spatio-Temporal Graph Convolutional Network for earthquake location and magnitude estimation. \citet{tan2024next} proposed SUGAR, a computer vision-based workflow that combines 3D U-Net neural networks with traditional earthquake location techniques to detect and locate seismic sources in complex sequences. \citet{shen2021array} developed ArrayConvNet, a two-stage CNN framework that directly detects and locates earthquakes from continuous multi-station waveform data without intermediate phase picking and association steps, achieving detection rates nearly 7 times higher than published catalogs when applied to Hawaiian seismic data. \citet{zhang2025multistation} developed MSLOC, which integrates station geometry into a 3D U-Net by constructing spatial data volumes from multi-station waveforms, achieving $\sim$5 km location accuracy.
However, in existing machine learning-based earthquake location methods, the accuracy of depth prediction remains low and significantly lags behind that of latitude and longitude. To date, no independent network has been developed that directly utilizes waveforms for depth determination. 

To enhance depth accuracy, particularly in regions with sparse station coverage, this study introduces a deep learning neural network to estimate earthquake depths by extracting waveform features from single stations. We propose VGGDepth, an independent depth prediction network capable of directly inferring depth from waveforms. The network architecture, derived from VGG16 \cite[]{simonyan2014very} in computer vision, accepts three-component waveforms from a single station as inputs and outputs corresponding depth, facilitating a direct mapping from waveforms to depth. We developed two types of deep learning models: individual models tailored to specific stations, and a generalized model applicable to any stations within a region.  For training and testing, we employed a high-precision earthquake catalog from the 2016-2017 Central Apennines, Italy earthquake sequence. The method was tested using catalogs from different time periods and varying numbers of stations in this region, demonstrating its potential for solving depth information during pseudo-sparse station conditions. This approach enables depth refinement for historical earthquakes recorded during periods of limited station coverage.

\section{Data and Method}

\subsection{VGGDepth: Single Station Model }

In this study, we developed VGGDepth, a neural network built upon the well-known VGG16 architecture in computer vision \cite[]{simonyan2014very}. VGGDepth maps single-station three-component waveforms to one-dimensional depth probability distributions. The detailed architecture of VGGDepth consists of five feature extraction combinatorial convolution modules and a fully connected (FC) feature translation module (Figure~\ref{fig: network}a). Each combinatorial convolution module includes a $3 \times 3$ convolution layer  ($\mathrm{ConV}$), a ReLU activation function, and a Max Pooling layer.  The inclusion of a batch normalization layer in each module further improves performance. The network depth and architecture were 
adjusted to suit the depth determination task. These convolution modules effectively extract depth-relevant features from the waveform data. Following the convolutional layers, the network includes three fully connected layers that translate the extracted depth features into depth probability distributions.  Notably, despite the existence of more advanced neural network architectures, for example, ResNet  \cite[]{he2016deep}, Transformer \cite[]{vaswani2017attention}, U-Net \cite[]{ronneberger2015u}, VGGDepth's structured design --- optimized for capturing depth-sensitive waveform features and efficiently mapping 2D seismic data to 1D depth distributions --- proves sufficient to meet the accuracy and reliability demands of earthquake depth determination in this study.

The input to the network is shaped as 1 (channel, means~single~station)$\times$3 (means three components) $\times$ 1024 (time samples), and the output is $\mathrm{batchsize \times 256~(depth~label)}$. We define a truncated time window as $[Ptime-2\,s, Stime+1.7*(Stime-Ptime)]$, which is then resampled from the original 100 Hz sampling rate to 20 Hz through interpolation. This time window contains the complete useful waveforms, including body waves, surface waves, and coda waves, while also helping to eliminate the effects of neighboring events and irrelevant noise. The waveform is zero-padded to a standardized input length of 1024 samples. The network outputs a one-dimensional Gaussian probability distribution representing earthquake depth, covering a monitoring range of 28 km with uniform mapping.

We trained the VGGDepth module using the PyTorch framework \cite[]{imambi2021pytorch}. For each station, we collected approximately 56,000 three-component waveform samples with a signal-to-noise ratio (SNR) above 5 dB. The SNR is computed as $20\times\log_{10}({\frac{RMS_{signal} }{ RMS_{noise}} )} $ in the 2-8 Hz frequency band (noise window: 3 seconds before P-wave arrival; signal window: P-wave arrival to $1.7\times (Stime-Ptime)$, matching the network input waveform ). Of these, 80\% ($\sim$45,000) were used for training and the remaining 20\% ($\sim$11,000) for validation. After testing several loss functions, we selected binary cross-entropy (BCE) as the optimal choice due to its consistently strong performance. The model was trained using hyperparameters optimized via grid search (learning rate = 0.00025, batch size = 256, 200 epochs), which were consistently applied to all single-station and regionally generalized models throughout this study. The learning rate was halved every 50 epochs. Additionally, training was stopped early if the depth error on the validation set increased for five consecutive epochs. Typically, it takes $\sim$1.6 hours to train a single-station model using a computer configured with a GeForce RTX 3090.

\begin{figure}[htpb]
\centering

\includegraphics[width=1\linewidth]{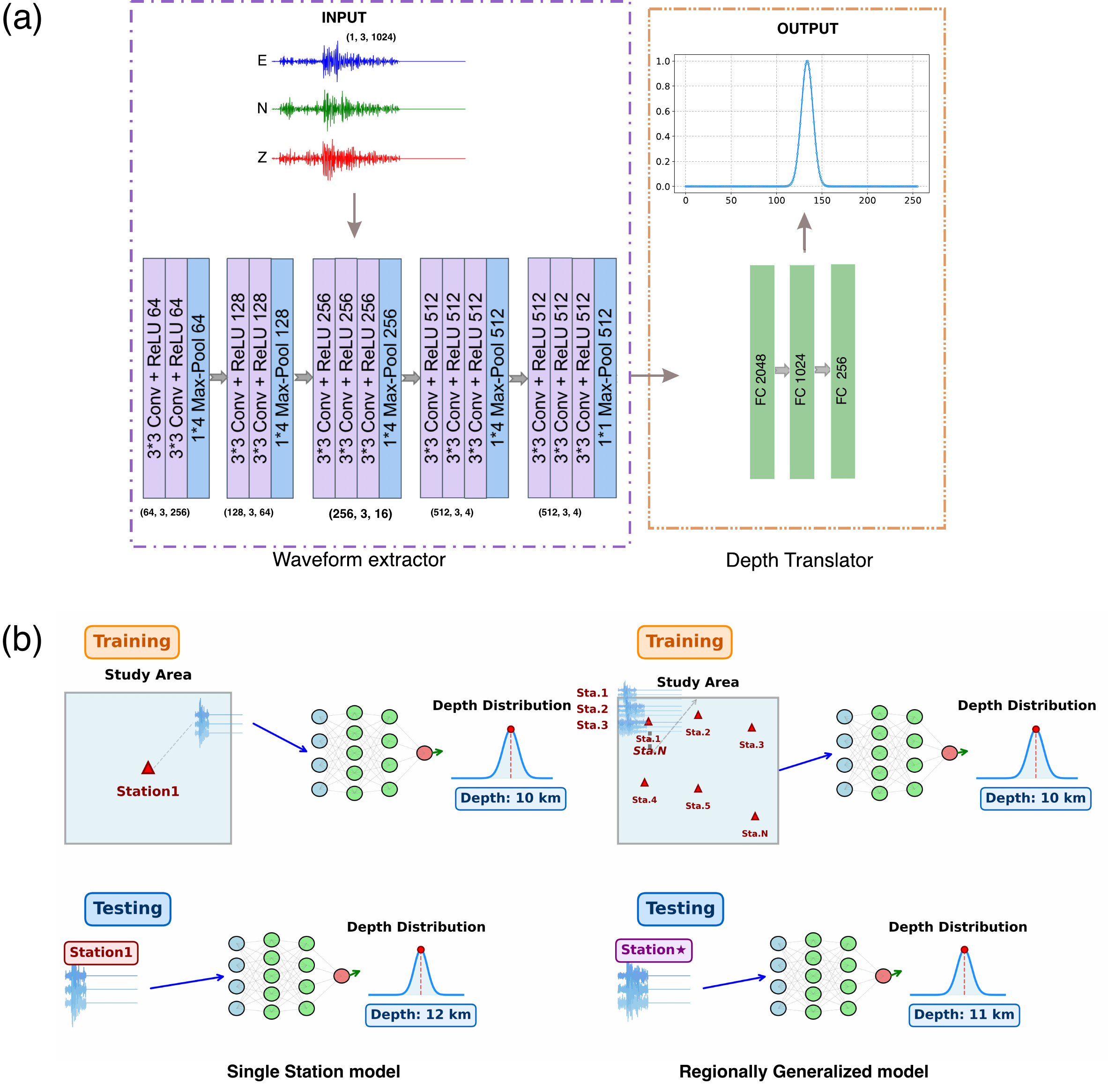}
\setlength{\abovecaptionskip}{2pt}  
\setlength{\belowcaptionskip}{1pt}
\caption{(a) Diagram showing the architecture of VGGDepth. The architecture consists of two integrated neural network modules: a feature extraction module and a deep feature translation module. The input to the network is a three-component waveform, which first passes through the feature extraction module --- a convolutional neural network-based structure. The resulting features are then fed into the deep feature translator, which is composed of a fully connected network. Finally, the network outputs the probability density function of the earthquake depth. (b) Comparison of single-station and regionally generalized earthquake depth prediction workflows. Left: Single-station model trained and tested on data from one station. Right: Regionally generalized model trained on data from multiple stations (Sta.1 to Sta.N) and tested across the network. Both models use the same architecture but differ in training strategy.
}\label{fig: network}
\end{figure}

\subsection{VGGDepth: Regionally Generalized Model}

We further introduced the concept of regional generalization, which involves training and testing the VGGDepth network using waveform data from multiple stations within a given region. This fundamental assumption is that earthquake waveforms with the same source depth recorded at different stations in the same region exhibit similar characteristics due to shared geological structures. The objective is to develop a model that generalizes across this region --- capable of processing waveform data from any stations within the region and improving depth estimation by averaging results across multiple stations. Moreover, this approach enables the use of a significantly larger number of waveform samples for network training. For instance, with 38 stations in the study region and a single station training set of 45,000, the theoretical training set size increases to $45,000 \times38$. The corresponding training time increased to 26 hours under the same computer configuration. After training, the resulting model follows a ``Many-to-One" paradigm, commonly used in computer vision, where waveform inputs from different stations for the same seismic event can be used to determine a single and consistent depth prediction. This generalized network allows the trained model to perform depth estimation at all existing stations across the region and potentially generalize to both legacy stations from the past and newly deployed stations in the future.

Both the single-station and regionally generalized models share the same neural network architecture and output format, but differ in their training strategy (Figure~\ref{fig: network}b).
The single-station model (Figure~\ref{fig: network}b, left) is trained and tested on waveforms from one specific station (e.g., Station1), learning station-specific features.
The regionally generalized model (Figure~\ref{fig: network}b, right) is trained on waveforms from all 38 stations in the study region. Each training sample still contains waveforms from a single station, but the training dataset includes examples from all stations. This enables the model to learn regional patterns that generalize across different station locations and site conditions. During testing, the trained model can predict earthquake depths for any stations within the study region.

\subsection{Dataset}

We demonstrated our method using the 2016-2017 Central Apennines, Italy earthquake sequence due to its abundant seismicity and high-precision earthquake catalog. On August 24, 2016, a moment magnitude  $M_w$ 6.0 earthquake struck the Central Apennines, Italy, causing almost 300 fatalities and the destruction of numerous ancient buildings. This was followed by two strong earthquakes ($M_w$ 6.1 and $M_w$ 6.6) near Visso and Norcia, forcing about 100,000 people to evacuate their homes and destroying many towns and cities \cite[]{dolce20182016}. The Central Apennines, situated at the convergent boundary between the Eurasian and African plates, represent one of the most seismically active regions in the Mediterranean due to complex interactions involving the northwestward subduction of the Adria microplate beneath the Apennine belt and subsequent post-collisional extensional tectonics. This area is characterized by an extensional stress regime, which has generated a network of high-angle normal fault systems aligned parallel to the orogenic axis \cite[]{chiarabba20092009}. These shallow-crustal faults (1-15 km depth) accumulate elastic strain from ongoing crustal thinning driven by slab break-off processes following Adria's subduction, leading to recurrent destructive earthquakes, as evidenced by historical events like the 1915 $M_w$ 6.7 earthquakes \cite[]{amoruso1998inversion}. The high seismic hazard in this region is another reason we selected it as the study area.

For model training and testing, we selected the high-precision earthquake catalog by \citet{tan2021machine}, which was constructed using a deep-learning phase picker followed by refined location procedures, ensuring reliable depth information for supervised learning. We defined a target region with a longitude range of 82 km and a latitude range of 112 km (Figure~\ref{fig: region}), which encompasses the majority of earthquake events in this area and offers relatively good seismic station coverage. Most epicentral distances fall within 100 km (Figure~\ref{fig: depth}). The depth range spans from 0 to 28 km, with the majority of events concentrated within 15 km --- providing a sufficient sample size for analysis (Figure~\ref{fig: depth}). We used seismic waveform data from both the permanent Network IV (triangles in Figure~\ref{fig: region}) and the temporary network YR (diamonds in Figure~\ref{fig: region}) in the study area. Waveform availability and quality were carefully verified before training.  
All waveforms were instrument-removed to account for different sensor types and were band-pass filtered between 2-8 Hz.  We systematically tested multiple frequency bands (1-10 Hz, 1-15 Hz, 2-8 Hz) and selected 2-8 Hz as it provides optimal noise suppression while preserving the essential seismic signal characteristics needed for depth estimation.
A truncated time window $[Ptime-2\,s, Stime+1.7*(Stime-Ptime)]$ with normalized amplitudes was used for both training and testing. 
For training (Section 3.1), we selected earthquake events that occurred between August 24, 2016, and December 31, 2016 --- a period encompassing three mainshocks (August 24, 2016, 
$ M_w$ 6.0; October 26, 2016, $ M_w$ 6.1; October 30, 2016, $ M_w$ 6.6) and a series of aftershocks --- with magnitudes ranging from $ M_w$ 0.5 to 6.6.  
For quality control, we excluded events with an SNR less than 5 dB. For performance evaluation (Section 3.2), we selected earthquake sequences that occurred in April-May 2017 (Figure S1), totaling 6600 aftershocks, with magnitudes ranging from 0.2 to 4.0.

\begin{figure}[htpb]
\centering\includegraphics[width=0.7\linewidth]{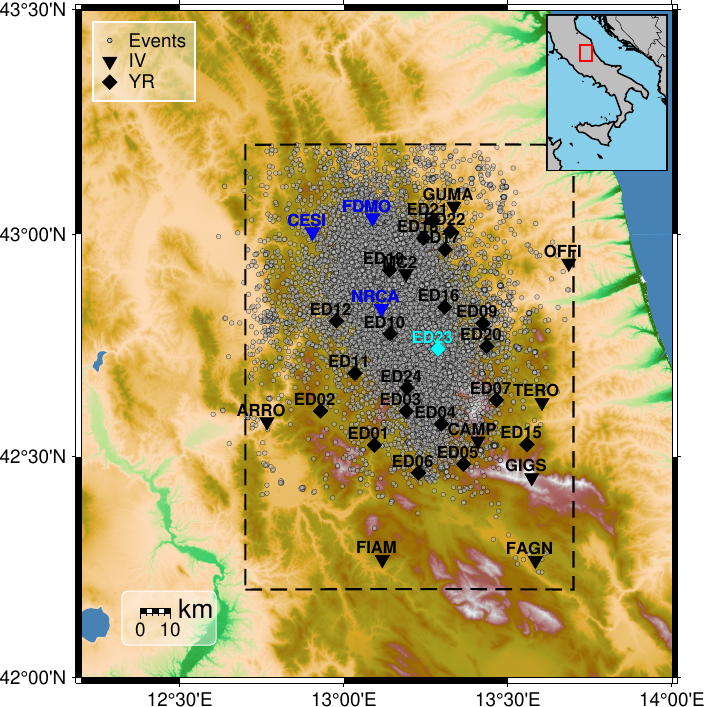}
\setlength{\abovecaptionskip}{3pt}  
\setlength{\belowcaptionskip}{0pt}
\caption{The study region with the distribution of earthquakes (gray dots) and stations including permanent network IV (triangles) and temporary network YR (diamonds). The black dashed box represents the study area. Three stations (NRCA, FDMO, CESI) were selected for evaluating single-station model performance. The station ED23 was selected to test the regionally generalized model for a pseudo-newly deployed station, as well as to evaluate the performance of transfer-learning.     }\label{fig: region}
\end{figure}

\begin{figure}[htpb]
\centering
 \noindent\includegraphics[width=1\textwidth]{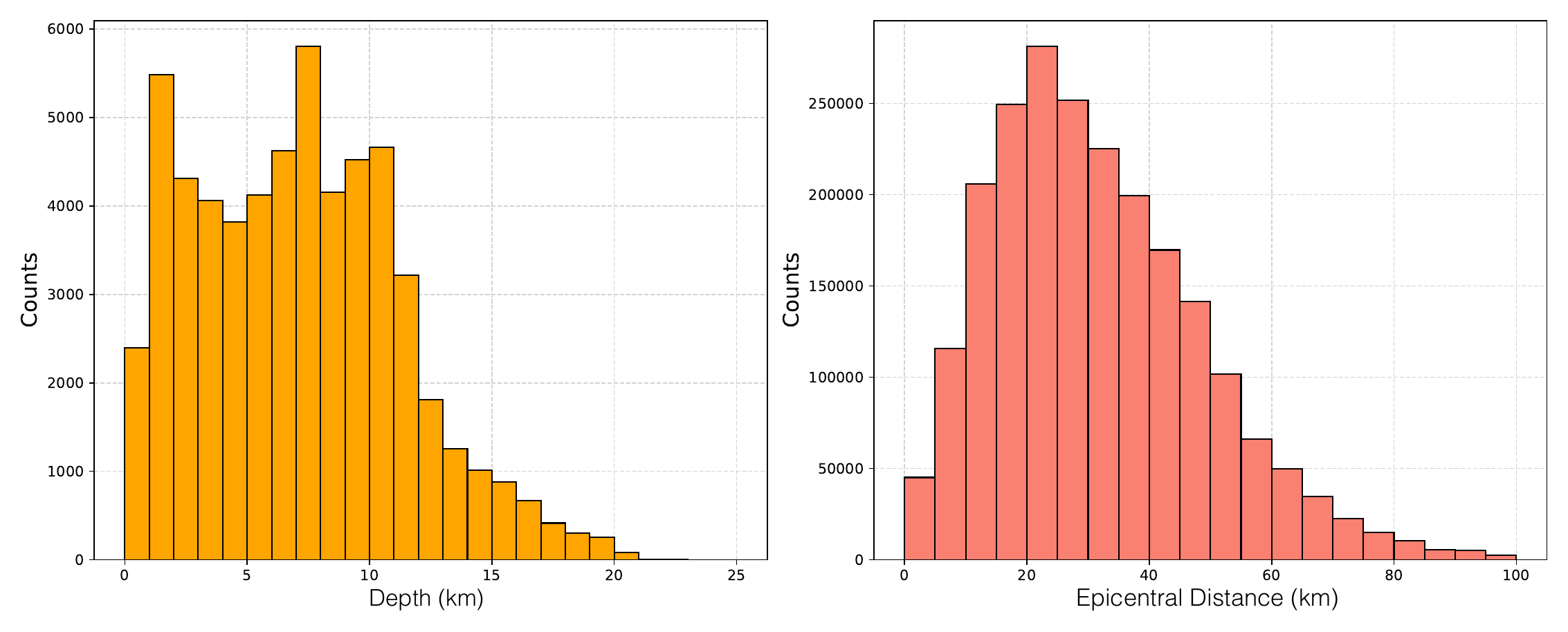}
 \setlength{\abovecaptionskip}{0pt}  
\setlength{\belowcaptionskip}{2pt}
\caption{Frequency distributions of earthquake depths (left panel) and epicentral distances (right panel). The depth statistics include all events, while the epicentral distance statistics are shown using station NRCA as an example. }
\label{fig: depth}
\end{figure}

\section{Results}

\subsection{Model Error Analysis}

During the training process, we adopted a real-time monitoring approach to analyze model errors using the validation set. The predicted depth is identified by evaluating the position of the maximum value of the depth prediction probability distribution (Figure S2). We then defined the depth error as the difference between the predicted depth and the labeled depth. This approach is effective and reliable, differing from those used in traditional deep learning training, which typically involves comparing the entire probability distributions of the prediction and the label. 

The model errors for the single station models and the regionally generalized model were evaluated across different epochs during their model training, respectively. As a demonstration, we selected three seismic stations for training and testing the single-station models. These three stations are separated by at least 20 km but are located within the same vicinity, making it likely that nearby earthquakes are recorded by all of them. This spatial configuration enables the use of multiple stations to improve depth estimation, as discussed in the following section.

The depth errors of all models stabilize and converge after 125 epochs (Figure~\ref{fig: error}), although substantial instability occurs during the first several dozen iterations. Overall, the single-station models exhibit smaller depth errors than the regionally generalized model. This is common in deep learning, where generalized models often compromise accuracy as more features are introduced. Among the three single-station models, strong variations are observed, which may be caused by differences in waveform data quality and local wave scattering.

\begin{figure}[htpb]
\centering\includegraphics[width=1\linewidth]{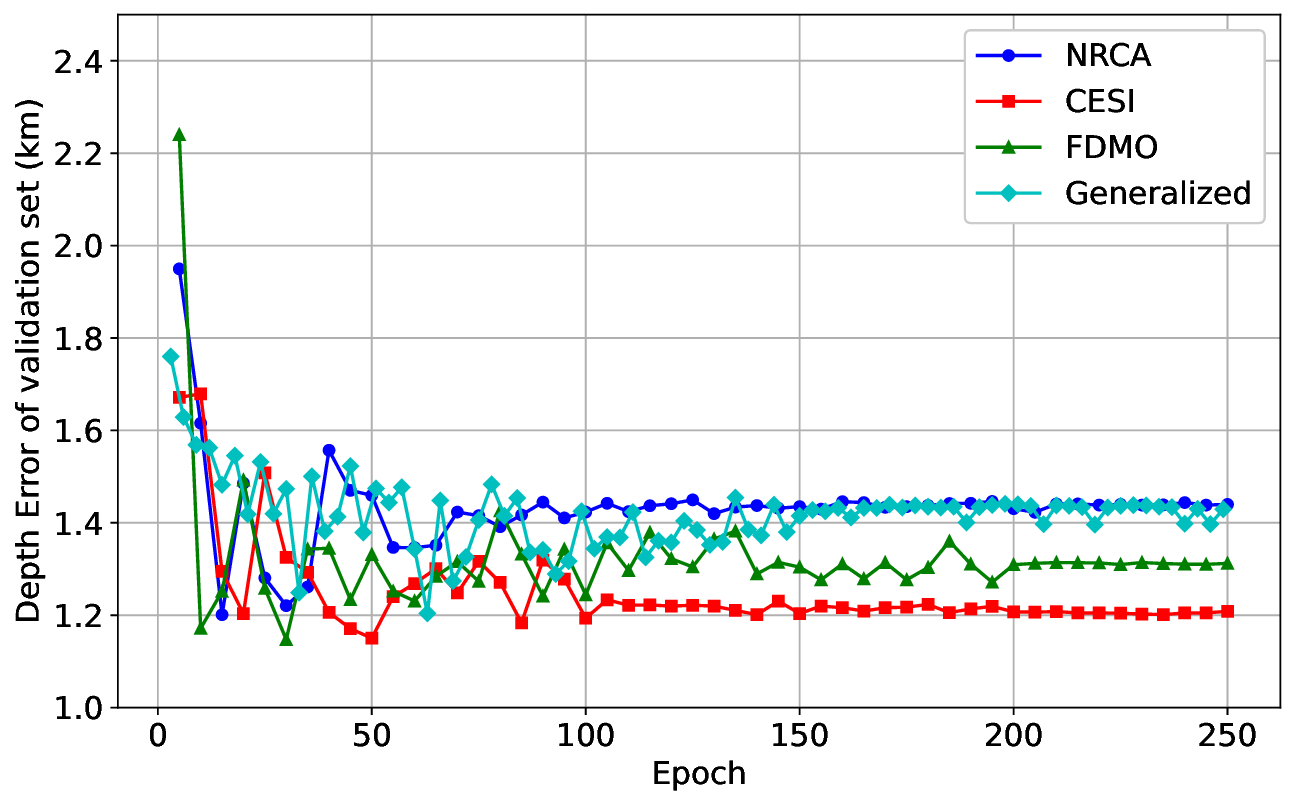}
\setlength{\abovecaptionskip}{2pt}  
\setlength{\belowcaptionskip}{4pt}
\caption{Depth error of the validation set as a function of training epochs. The blue, red, and green colored lines denote three individual single-station models trained on three different stations in the study area (i.e., NRCA, CESI, and FDMO), while the light blue line represents the regionally generalized model for all stations in the region. 
}\label{fig: error}
\end{figure}

\subsection{ Model Performance Testing  }

We evaluated the depth prediction performance of the single-station models and the regionally generalized model using earthquakes that occurred approximately half a year after the training events in the Central Apennines region, specifically from April to May 2017. This testing set includes a total of 6600 aftershocks, with magnitudes ranging from 0.2 to 4. For single-station model evaluation, each model was tested on its corresponding seismic station. For the evaluation of the regionally generalized model, the same model was applied to all three stations. Consistent with the training process, we estimated the predicted depth as the value corresponding to the maximum of the depth prediction probability distribution, provided that the probability exceeds a defined threshold. A higher threshold results in fewer solvable events but more accurate predictions, and vice versa.

To comprehensively evaluate the model performance and determine appropriate probability thresholds, we computed confusion matrices, precision-recall curves, and F1-scores for both models (Figure~\ref{fig: Normal_AUCPR} and Figure~\ref{fig: Generalized_AUCPR}), with $\pm$1 km depth error and the probability threshold serving as the criteria. The area under the precision-recall curve (AUC-PR) values range from 0.801 to 0.833 for single-station models and 0.809 to 0.863 for the regionally generalized model, indicating robust performance for both approaches in balancing precision and recall. The F1-score analysis (Figure~\ref{fig: Normal_AUCPR}g-i and Figure~\ref{fig: Generalized_AUCPR}g-i) shows that the optimal probability thresholds for maximum F1-scores are approximately 0.956 for single-station models and 0.987 for the regionally generalized model. Based on these results, we select probability thresholds of 0.95 and 0.99 for the single-station and 
regionally generalized models, respectively, throughout this study. These 
thresholds are close to the optimal values while providing a balance between prediction accuracy and event coverage.

With these selected probability thresholds (0.95 for single-station and 0.99 for regionally generalized), the single-station models (Figure~\ref{fig: Normal}) achieve depth errors ranging from 0.87 km to 0.98 km across the three stations, with approximately 71\%-85\% of the events yielding valid depth estimates from their individual stations. The regionally generalized model (Figure~\ref{fig: Generalized}) also performs well, producing slightly smaller errors of 0.93 km, 0.85 km, and 0.81 km at stations NRCA, CESI, and FDMO, respectively, compared to the single-station models. Training with all stations enables the network to achieve a more complete convergence, thereby allowing the use of a higher threshold for predictions. However, the number of solvable events consistently decreases across all stations (69\%-72\%). At these thresholds, the single-station models achieve F1-scores of 0.778-0.835 across the three stations, while the regionally generalized model yields comparable F1-scores of 0.775-0.805.

We evaluated depth resolution using normalized recovery matrices, where each cell (X, Y) represents the proportion of events with true depth Y that are predicted at depth X, normalized by the total number of events at that true depth. The normalized recovery matrices (Figure~\ref{fig: Normal}g-i and Figure~\ref{fig: Generalized}g-i) demonstrate reliable depth resolution for both models, with diagonal values exceeding 0.5-0.7 for most depth bins. For very deep events, there is a slight off-diagonal shift toward shallower predicted depths, indicating systematic underestimation. This bias is attributed to the limited number of training samples at these extreme depths, which is characteristic of the natural seismicity distribution in the Central Apennines, where deep earthquakes ($>$14 km) are less frequent. Despite this limitation, the model maintains reasonable accuracy with diagonal values exceeding 0.5 for most depth bins across the entire 0-12 km range.

\begin{figure}[htpb]
\centering\includegraphics[width=1\linewidth]{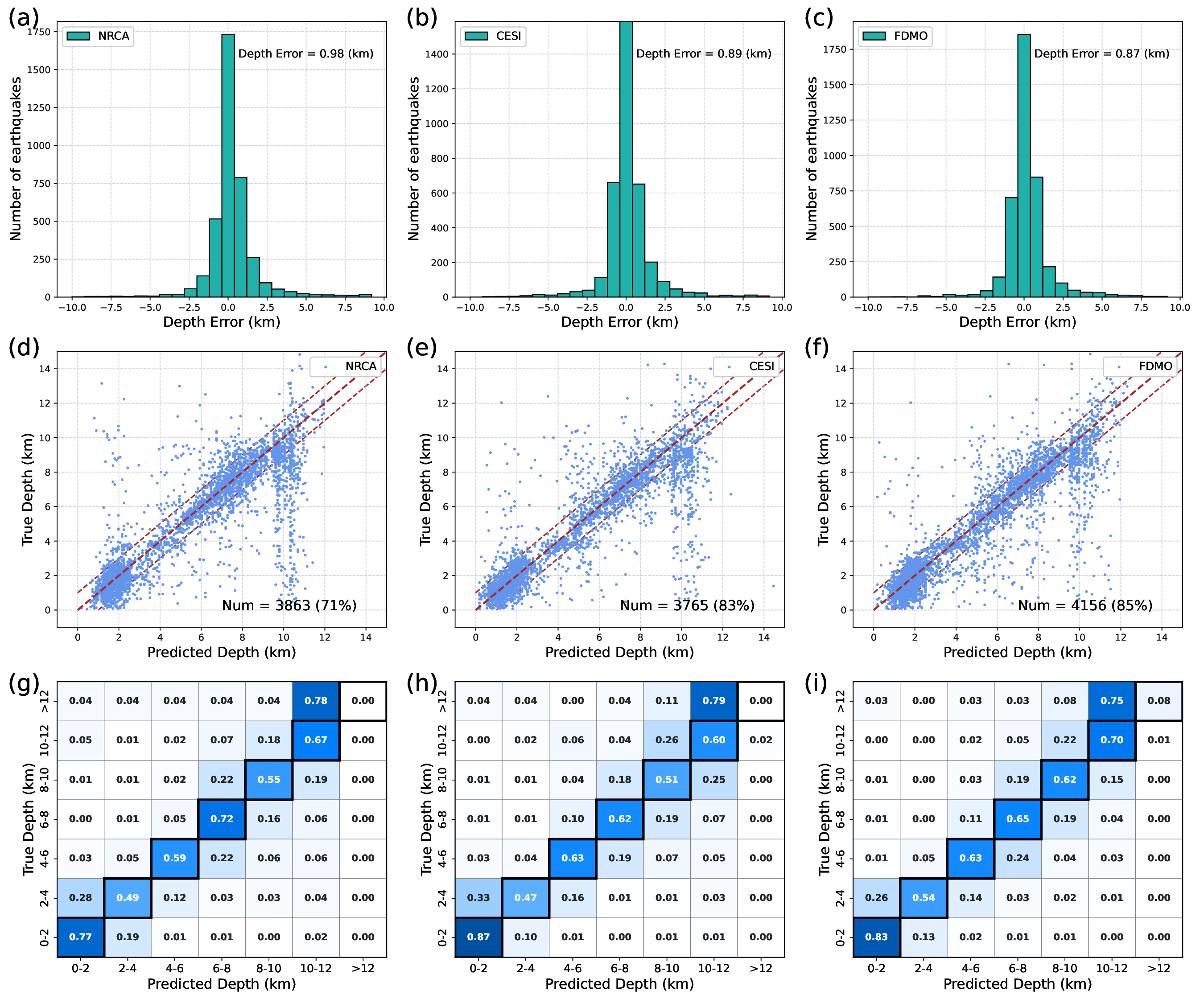}
\setlength{\abovecaptionskip}{2pt}  
\setlength{\belowcaptionskip}{2pt}
\caption{Depth prediction performance on the testing sets of the three seismic stations (stations NRCA, CESI, and FDMO). The top panels (a-c) show the depth error statistics, and the middle panels (d-f) display the depth comparisons between the predictions and their labels for individual events (blue dots). The mean depth errors and the percentages of the solvable events are indicated in the text within the figures. The bottom panels (g-i) present the normalized recovery matrices, where each cell (X, Y) represents the proportion of events with true depth Y that are predicted at depth X, normalized by the total number of events at that true depth. The diagonal elements indicate accurate depth recovery, with values exceeding 0.6 for most depth bins.   }\label{fig: Normal}
\end{figure}

\begin{figure}[htpb]
\centering\includegraphics[width=1\linewidth]{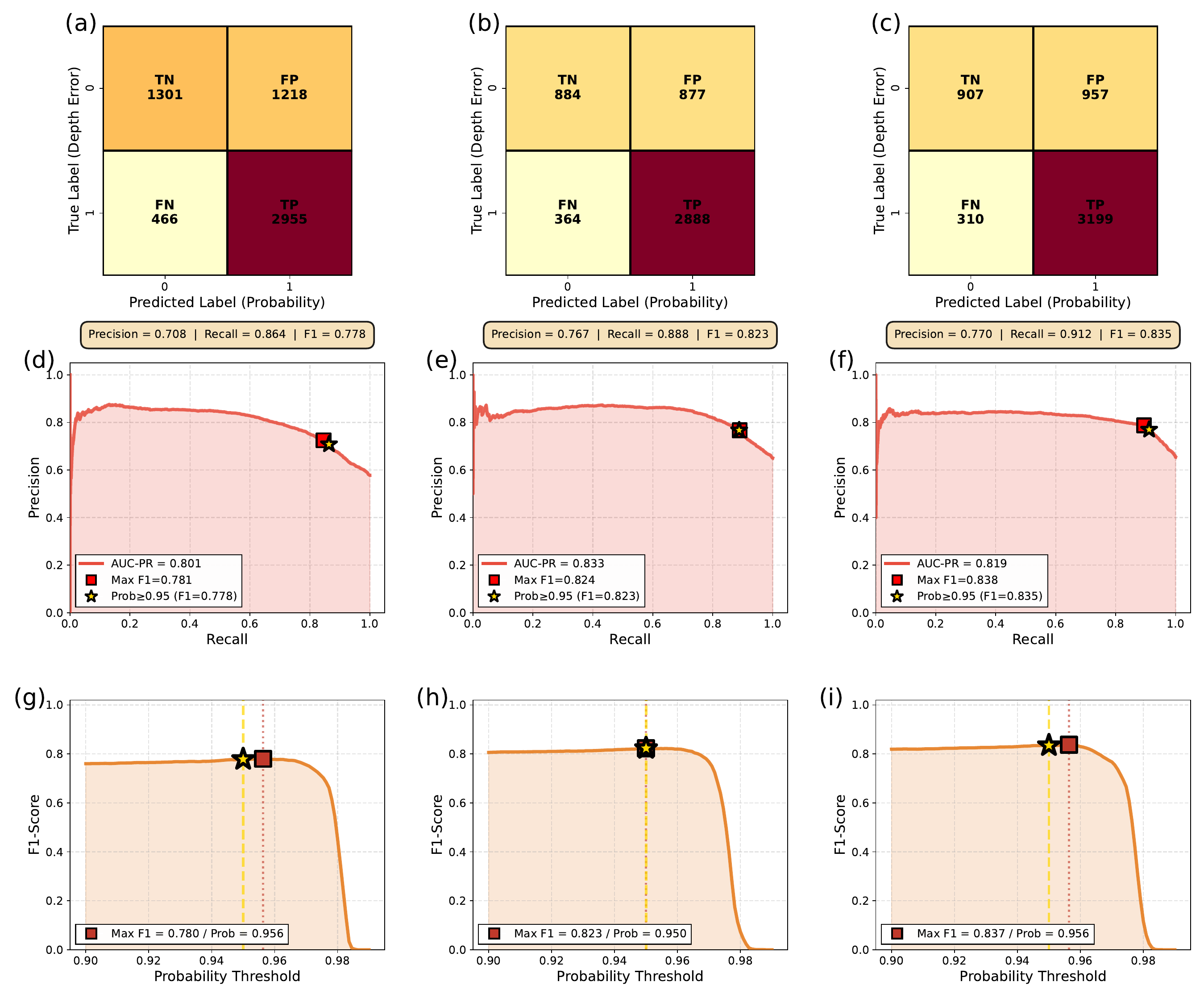}
\setlength{\abovecaptionskip}{2pt}  
\setlength{\belowcaptionskip}{2pt}
\caption{Performance evaluation of the single-station models at three representative stations (NRCA, CESI, and FDMO) using confusion matrices, precision-recall analysis, and F1-score optimization. (a-c) Confusion matrices showing true negatives (TN), false positives (FP), false negatives (FN), and true positives (TP) with a depth error threshold of $\pm$1 km and probability threshold of 0.95, along with precision, recall, and F1-scores. (d-f) Precision-recall (PR) curves with area under the curve (AUC-PR) values, showing the trade-off between precision and recall across different probability thresholds. The square markers indicate the maximum F1-score, and the star markers show the performance at the probability threshold of 0.95. (g-i) F1-score as a function of probability threshold, with vertical lines indicating the threshold for maximum F1-score (dashed darkred line) and the selected threshold of 0.95 (yellow line).   }\label{fig: Normal_AUCPR}
\end{figure}

\begin{figure}[htpb]
\centering\includegraphics[width=1\linewidth]{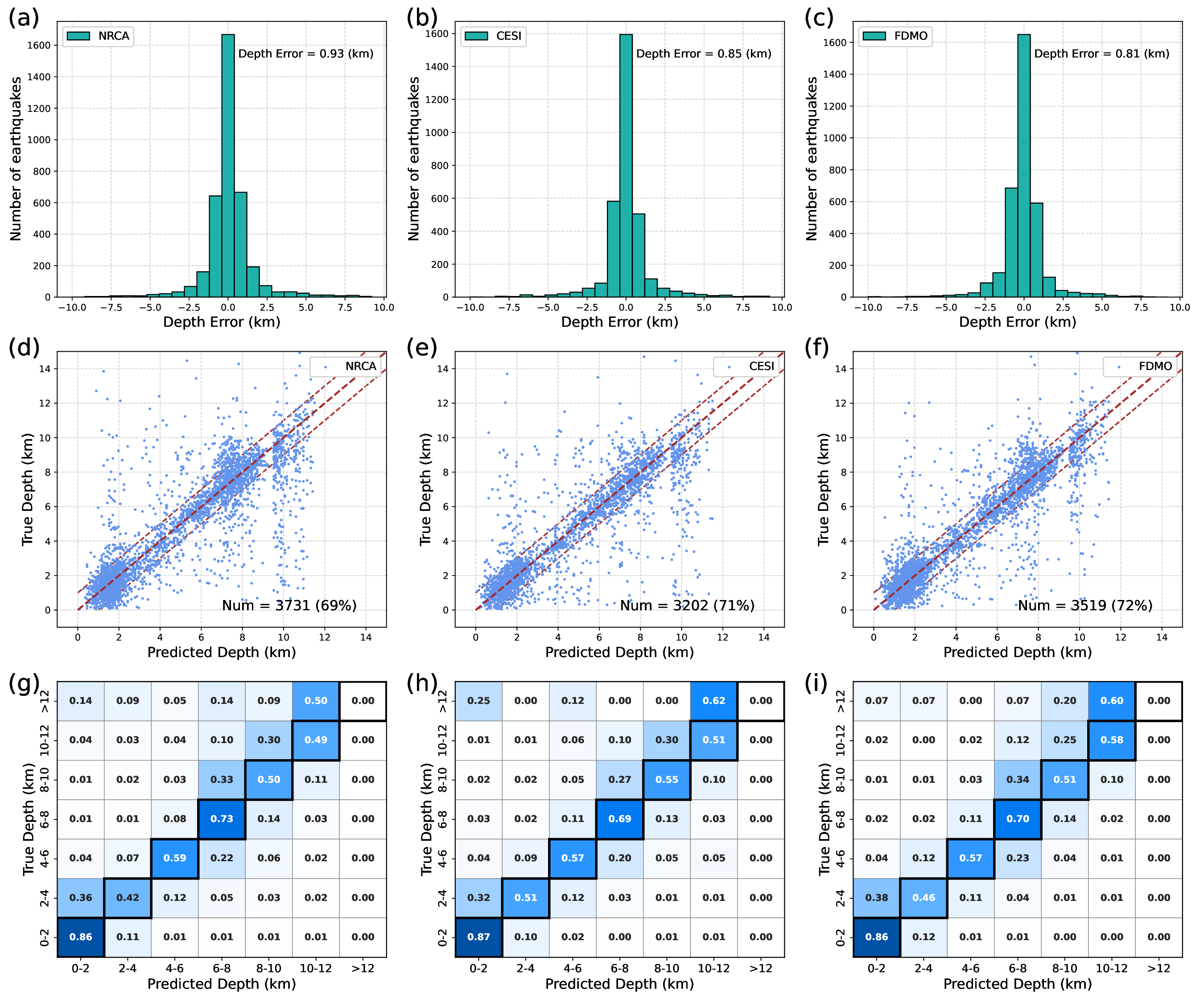}
\setlength{\abovecaptionskip}{2pt}  
\setlength{\belowcaptionskip}{0pt}
\caption{Same as Figure~\ref{fig: Normal}, except that the depth predictions for the three stations are obtained using the regionally generalized model. }\label{fig: Generalized}
\end{figure}

\begin{figure}[htpb]
\centering\includegraphics[width=1\linewidth]{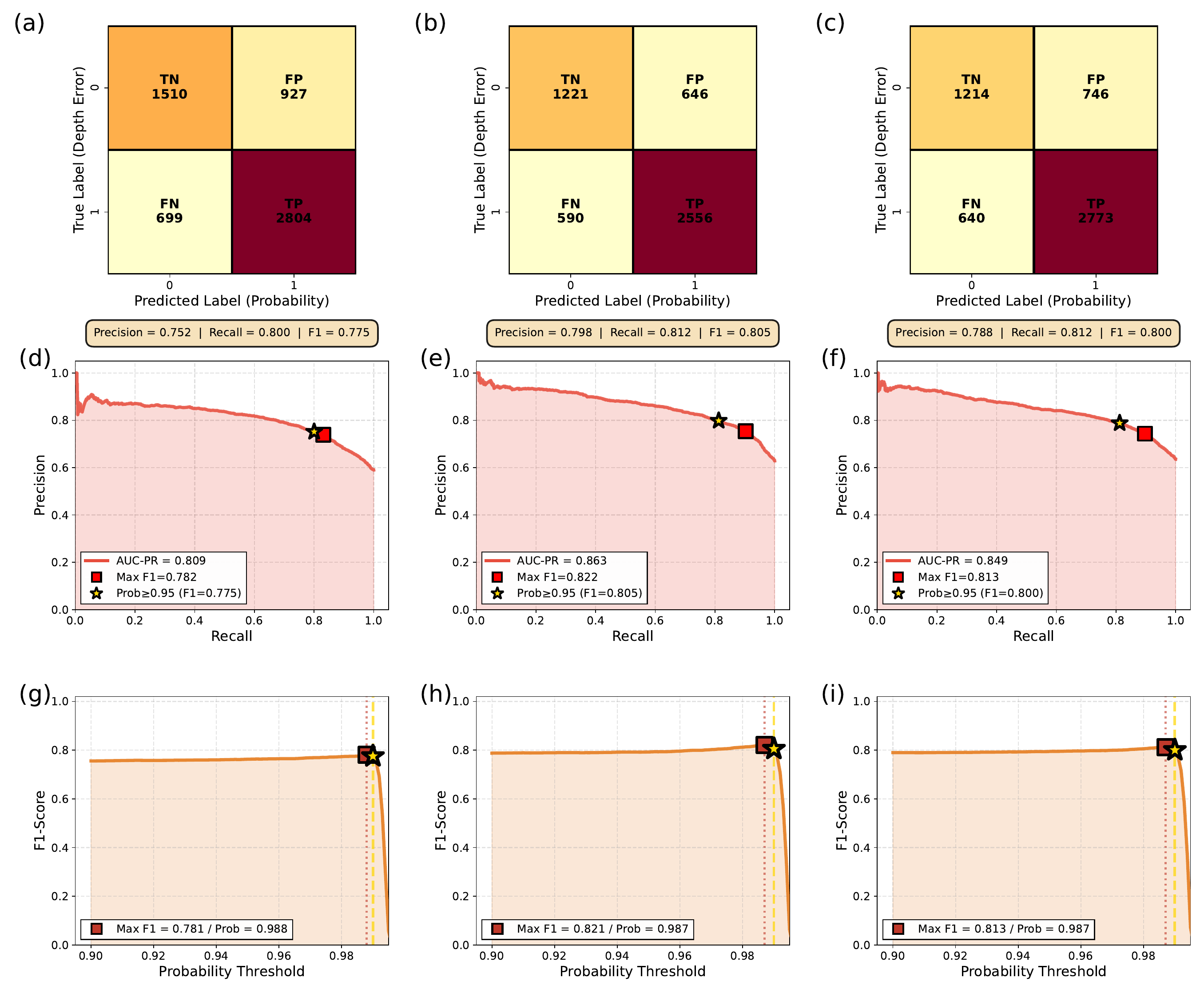}
\setlength{\abovecaptionskip}{2pt}  
\setlength{\belowcaptionskip}{0pt}
\caption{Same as Figure~\ref{fig: Normal_AUCPR}, except that the depth predictions for the three stations are obtained using the regionally generalized model. }\label{fig: Generalized_AUCPR}
\end{figure}

\subsection{Performance of Three-Station Averaged Depth Prediction Based on Single-Station Models and the Regionally Generalized Model  }


The single-station models perform excellently at each of the three stations individually, while the depth prediction of an earthquake can be further improved by averaging the results from all three stations. We selected 3960 earthquakes that were recorded by stations NRCA, CESI, and FDMO with eligible waveform quality (i.e., SNR $>$ 5 dB) to evaluate three-station averaged depth predictions using their corresponding single-station models and the regionally generalized model. Our analysis reveals that both the single-station models and the regionally generalized model significantly outperform individual station predictions, reducing depth errors substantially --- from $\sim$0.9 km to 0.62 km for single-station models and to 0.78 km for the regionally generalized model --- and exhibiting far fewer outliers (Figure~\ref{fig: ave}). The distribution of prediction standard deviations (the inset histograms in Figure~\ref{fig: ave}b,e) across the three stations, with most events having standard deviations below 1.5 km, indicates good inter-station consistency. The normalized recovery matrices (Figure~\ref{fig: ave}c,f) demonstrate strong diagonal patterns with values exceeding 0.6-0.8 for most depth bins, indicating reliable depth recovery across the 0-12 km range.

To optimize the performance of multi-station averaging, we analyzed the confusion matrices, precision-recall curves, and F1-scores for both models (Figure~\ref{fig: ave_AUCPR}). The AUC-PR values are 0.919 for single-station models and 0.922 for the regionally generalized model, indicating robust performance. The F1-score analysis (Figure~\ref{fig: ave_AUCPR}c,f) reveals that the optimal probability thresholds are approximately 0.901 for single-station models and 0.9 for the regionally generalized model. Importantly, multi-station averaging enables us to relax the probability threshold while maintaining high precision, as prediction uncertainties from individual stations are reduced through averaging. For the regionally generalized model, the F1-score curve remains relatively stable for probability thresholds above 0.90, allowing us to select a lower threshold of 0.92 for multi-station averaging tests. By using this relaxed threshold of 0.92, the regionally generalized model achieves a substantial improvement in recovery rate from 71\% to 92\% while maintaining comparable precision.

Although waveform characteristics may vary across the three stations, depth information is consistently embedded in their recordings. However, since the final depth estimate is averaged across all three stations, each station must exceed the defined prediction probability threshold. For multi-station averaging, we use a relaxed threshold of 0.92 for both models, compared to 0.95 and 0.99 in single-station predictions.
With this relaxed threshold, the single-station models maintain a recovery rate of 78\% (versus 79\% average in single-station predictions) while reducing depth errors from $\sim$0.9 km to 0.62 km (Figure~\ref{fig: ave}). Multi-station averaging thus enables us to relax the threshold while improving precision --- a balance unattainable with stricter thresholds that would sacrifice recovery rates.
The regionally generalized model shows even greater improvement. Using the same 0.92 threshold, recovery rate increases from 71\% to 92\% while depth error decreases from 0.81 km to 0.78 km (Figure~\ref{fig: ave}). Training on multi-station data enables more complete convergence, allowing the relaxed threshold to maintain high recovery rates while multi-station averaging preserves precision. Incorporating additional stations could further enhance prediction reliability and accuracy for both approaches, as discussed in the next section.

\begin{figure}[htpb]
\centering
\includegraphics[width=1\textwidth]{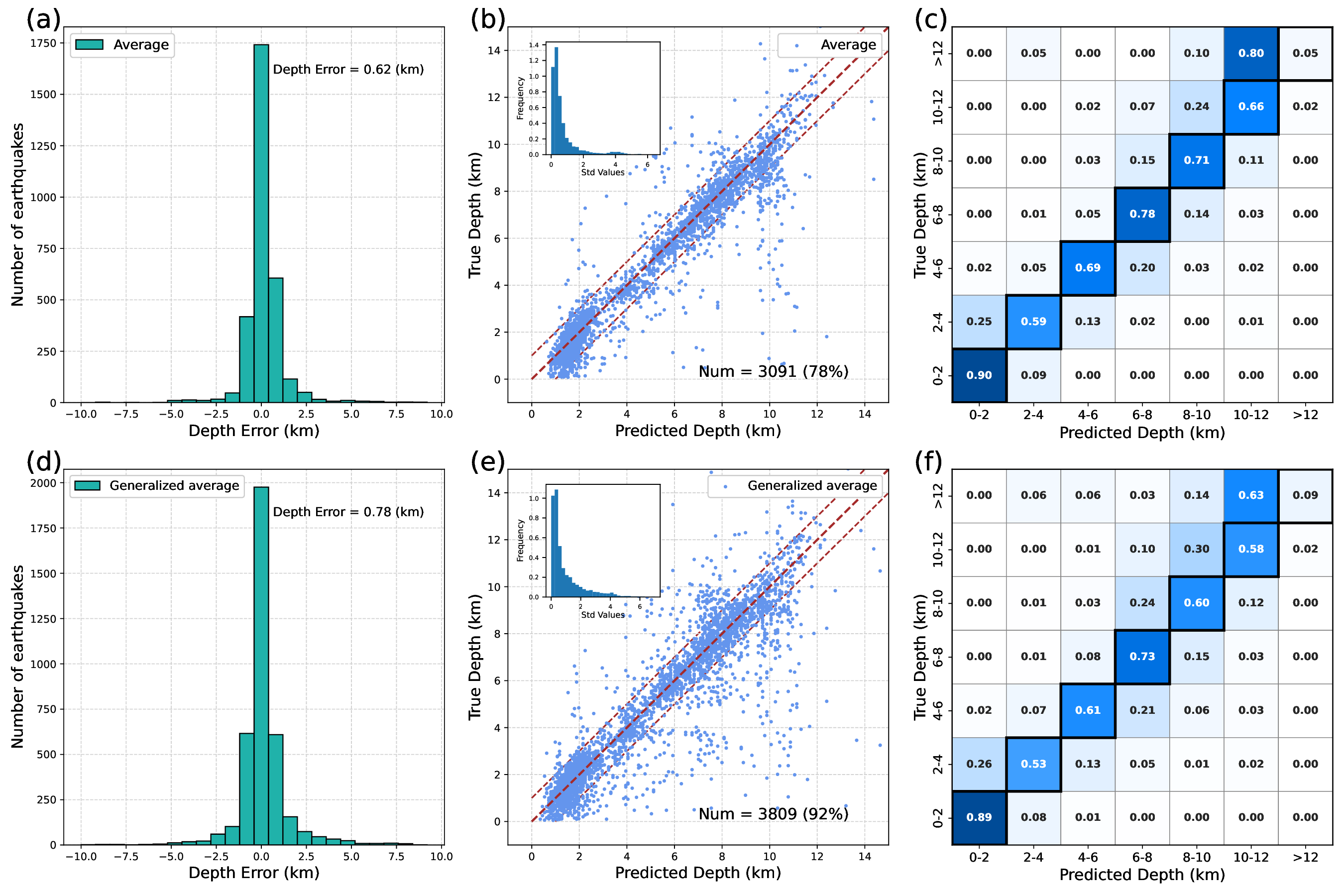}
\setlength{\abovecaptionskip}{2pt}  
\setlength{\belowcaptionskip}{2pt}
\caption{Depth prediction performance using the average depth obtained from three stations (NRCA, CESI, and FDMO) based on the single-station models (a-c) and the regionally generalized model (d-f). Panels (a-c) represent the depth averages from the three stations using their corresponding single-station models, while panels (d-f) show the corresponding averages using the regionally generalized model. The left panels (a, d) show the depth error statistics, the middle panels (b, e) display the depth comparisons between the predictions and their labels for individual events (blue dots), with inset histograms showing the 
distribution of prediction standard deviations across the three stations for each event. The right panels (c, f) present the normalized recovery matrices. The mean depth errors and the percentages of the solvable events are indicated in the text within the figures.  }
\label{fig: ave}
\end{figure}

\begin{figure}[htpb]
\centering
\includegraphics[width=1\textwidth]{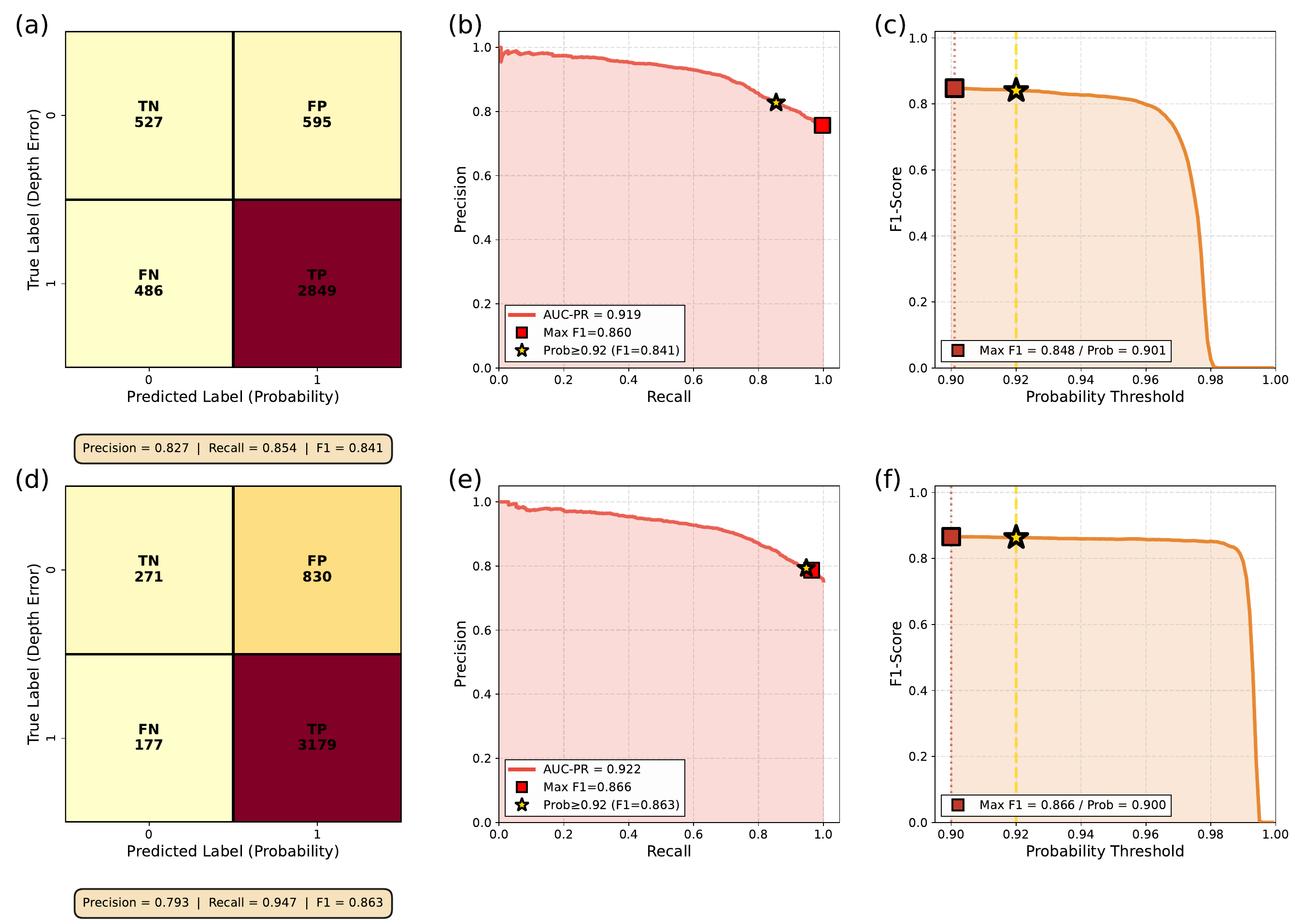}
\setlength{\abovecaptionskip}{2pt}  
\setlength{\belowcaptionskip}{2pt}
\caption{Same as Figure~\ref{fig: Normal_AUCPR}, except that the depth predictions are obtained by averaging results from three stations (NRCA, CESI, and FDMO) using single-station models (a-c) and the regionally generalized model (d-f). The probability threshold of 0.92 is used for both models in multi-station averaging. }
\label{fig: ave_AUCPR}
\end{figure}

\subsection{Performance of Regional-Station Averaged Prediction Based on the Regionally Generalized Model  }


Similar to the three-station averaged depth prediction, we evaluated the depth prediction performance using the average of regional stations based on the regionally generalized model (Figure~\ref{fig: ave_com_epi12}). Instead of using all available stations, we limited the analysis to the 12 nearest ones, as earthquakes are most likely to be recorded by nearby stations. To simulate realistic conditions, we extracted waveforms from these 12 stations without considering their waveform quality. For low-magnitude events, it is possible that some stations do not record effective waveforms or only record low-SNR data, which may result in depth prediction probabilities falling below the threshold --- ultimately preventing depth estimation for some events. If the prediction probability exceeds our threshold of 0.92, we count it as effective. Accordingly, we considered different numbers of effective station counts in our analysis, including more than 3, 6, 8, and 10 stations (Figure~\ref{fig: ave_com_epi12}a-l).
As the number of eligible stations increases, the depth errors systematically decrease. Specifically, increasing the minimum station count from 3 to 10 reduces the depth error from 0.94 km to 0.59 km (Figure~\ref{fig: ave_com_epi12}a,d,g,j). Most events have multi-station prediction standard deviations within 1.5 km (as shown in the inset histograms in Figure~\ref{fig: ave_com_epi12}b,e,h,k), similar to the three-station average test case, demonstrating good inter-station consistency.
 The normalized recovery matrices (Figure~\ref{fig: ave_com_epi12}c,f,i,l) demonstrate progressively stronger diagonal patterns with increasing station counts, with diagonal values reaching 0.7-0.8 for most depth bins when using 10 or more stations. However, the event recovery rate also declines, dropping from 99\% to 85\% (Figure~\ref{fig: ave_com_epi12}b,e,h,k). Thus, there is a trade-off between depth accuracy and event recovery rate.

To further evaluate the trade-off between depth accuracy and event recovery, we analyzed the confusion matrices and precision-recall curves for different station count requirements (Figure~\ref{fig: ave_com_epi12_AUCPR}). With a fixed probability threshold of 0.92 (suitable for multi-station average work), the confusion matrices (Figure~\ref{fig: ave_com_epi12_AUCPR}a-d) show that increasing the minimum station count from 3 to 10 improves precision from 0.776 to 0.864 while reducing recall from 0.998 to 0.893, resulting in F1-scores ranging from 0.873 to 0.893. The precision-recall curve (Figure~\ref{fig: ave_com_epi12_AUCPR}e) demonstrates that station counts of 6-8 achieve well-balanced F1-scores exceeding 0.885, making them optimal choices for most applications. The AUC-PR value of 0.903 indicates robust performance across different station count requirements. As the minimum station count increases from 3 to 10, MAE systematically decreases, clearly demonstrating the continuous trade-off between depth accuracy and event coverage (Figure~\ref{fig: ave_com_epi12_AUCPR}f).

\begin{figure}[htpb]
\centering
\includegraphics[width=0.9\textwidth]{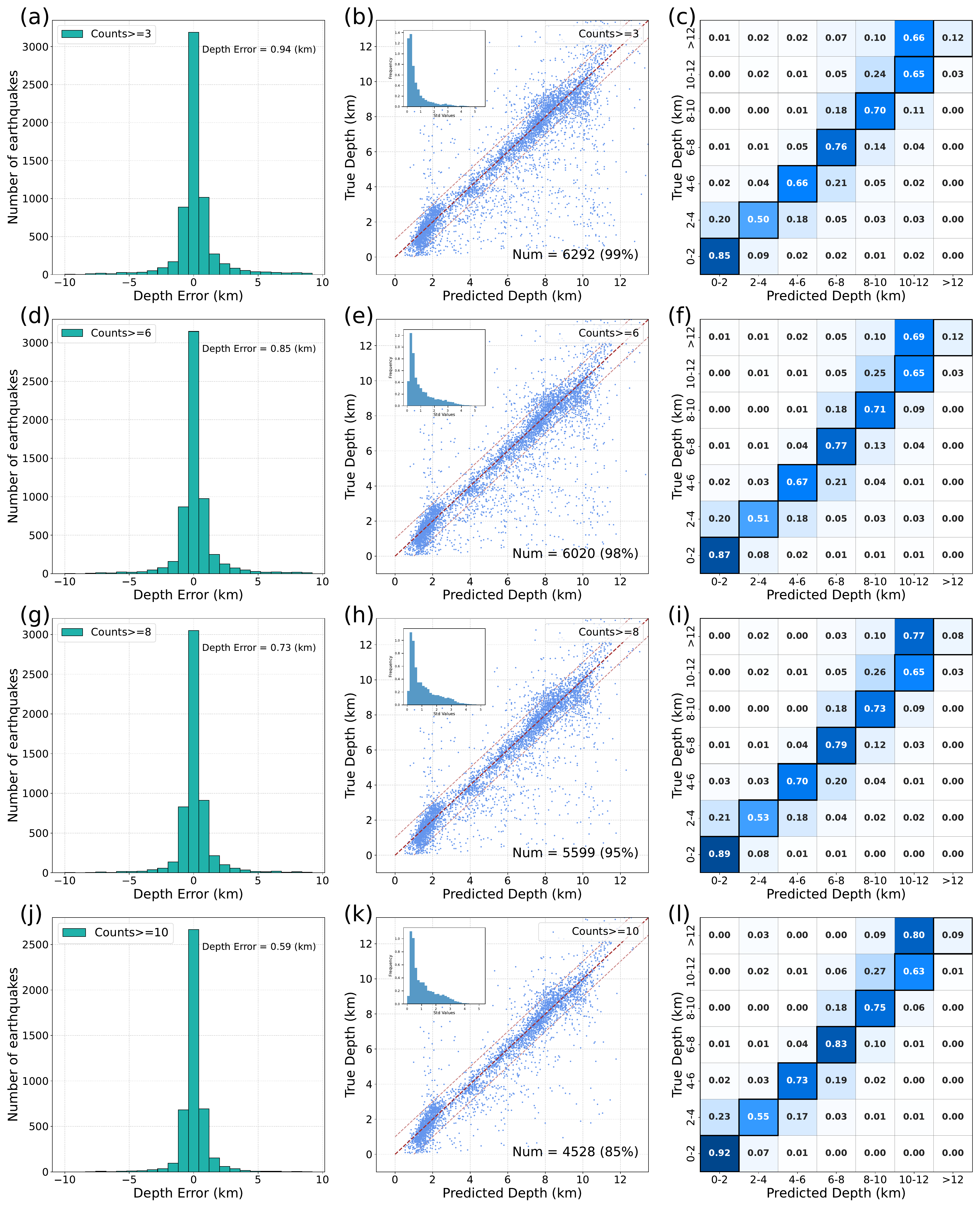}
\setlength{\abovecaptionskip}{2pt}  
\setlength{\belowcaptionskip}{2pt}
\caption{Depth prediction performance using the average of the 12 closest stations based on the regionally generalized model. Different numbers of eligible stations were analyzed, including cases with more than 3 (a-c), 6 (d-f), 8 (g-i), and 10 (j-l) stations. Data from a station is counted as eligible if it satisfies the prediction probability threshold and the SNR. The left panels (a, d, g, j) show the depth error distributions, the middle panels (b, e, h, k) display the depth comparisons between the predictions and their labels for individual events (blue dots), with inset histograms showing the distribution of prediction standard deviations across multiple stations for each event, and the right panels (c, f, i, l) present the normalized recovery matrices. The mean depth errors and the percentages of the solvable events are indicated in the text within the figures. }
\label{fig: ave_com_epi12}
\end{figure}

\begin{figure}[htpb]
\centering
\includegraphics[width=0.9\textwidth]{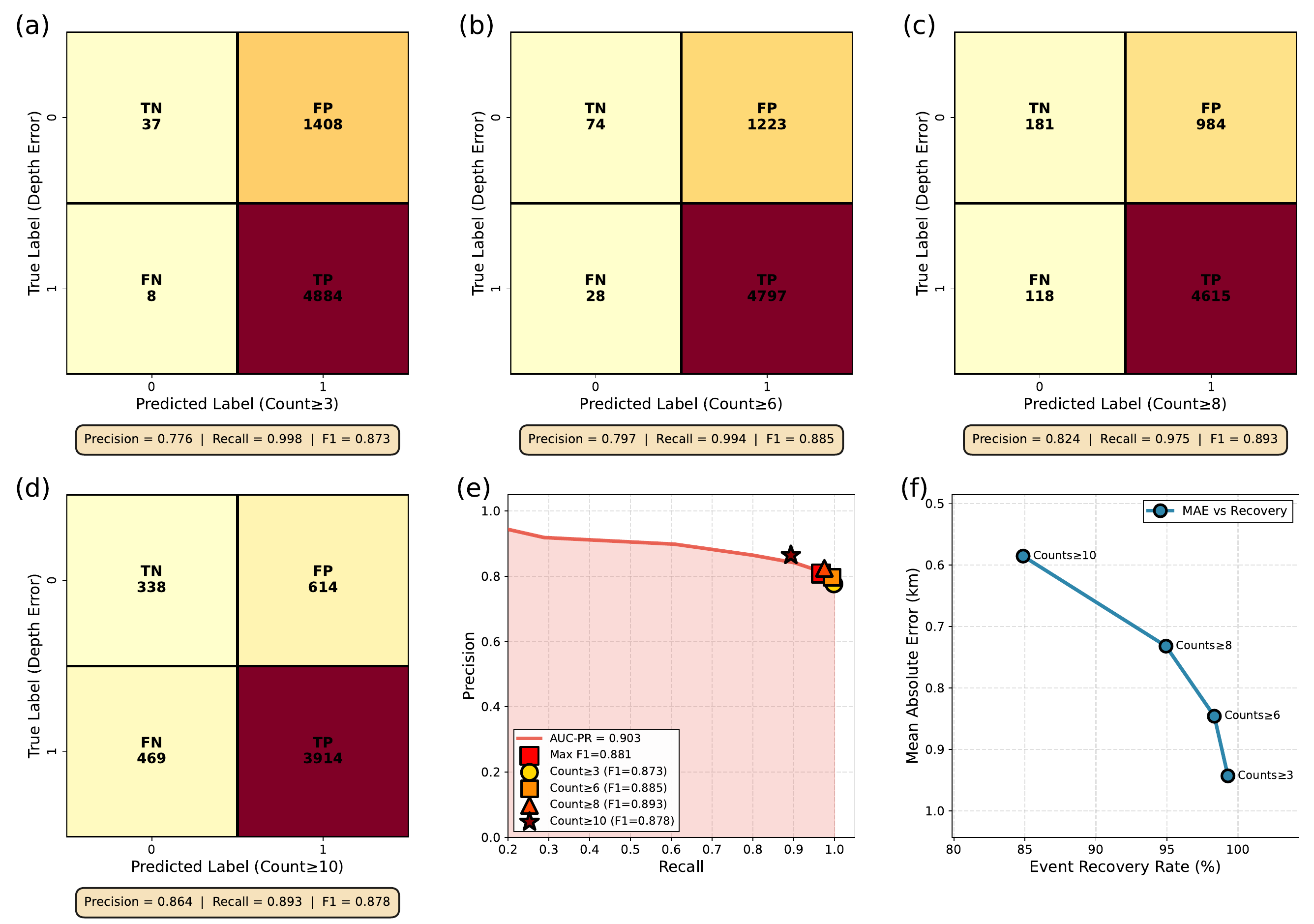}
\setlength{\abovecaptionskip}{2pt}  
\setlength{\belowcaptionskip}{2pt}
\caption{Confusion matrices (a-d) and precision-recall curve (e) for multi-station averaged depth predictions with different minimum station count requirements (3, 6, 8, and 10 stations). Panel (f) shows the trade-off between mean absolute error and event recovery rate. All analyses use a fixed probability threshold of 0.92.  }
\label{fig: ave_com_epi12_AUCPR}
\end{figure}

\subsection{Performance of the Regionally Generalized Model for Lack of Nearby Stations}


To analyze the relationship between depth errors and epicentral distance, we applied the regionally generalized model to 6,600 test earthquakes recorded at all 38 individual stations. To better visualize and compare performance as a function of epicentral distance, we ranked the 12 closest stations for each event and analyzed their depth errors individually (Figure~\ref{fig: closet_epi}a-l). The epicentral distances span a wide range, with the closest stations (1st-3rd) predominantly concentrated within 0-15 km, while more distant stations (9th-12th) extend beyond 30 km. The mean absolute error (MAE) and binned statistics (Figure~\ref{fig: closet_epi}m) demonstrate that depth errors remain stable across different epicentral distances, with minimal systematic bias. Depth errors and their standard deviations exhibit very minor variation for distances less than 30 km, while slightly larger variations occur at greater distances, likely due to reduced sample sizes caused by lower SNR or prediction probabilities falling below the threshold. Overall, these results demonstrate that the regionally generalized model maintains consistent depth prediction accuracy regardless of epicentral distance, highlighting its weak dependence on station proximity.

\begin{figure}[htpb]
\centering
\includegraphics[width=1\textwidth]{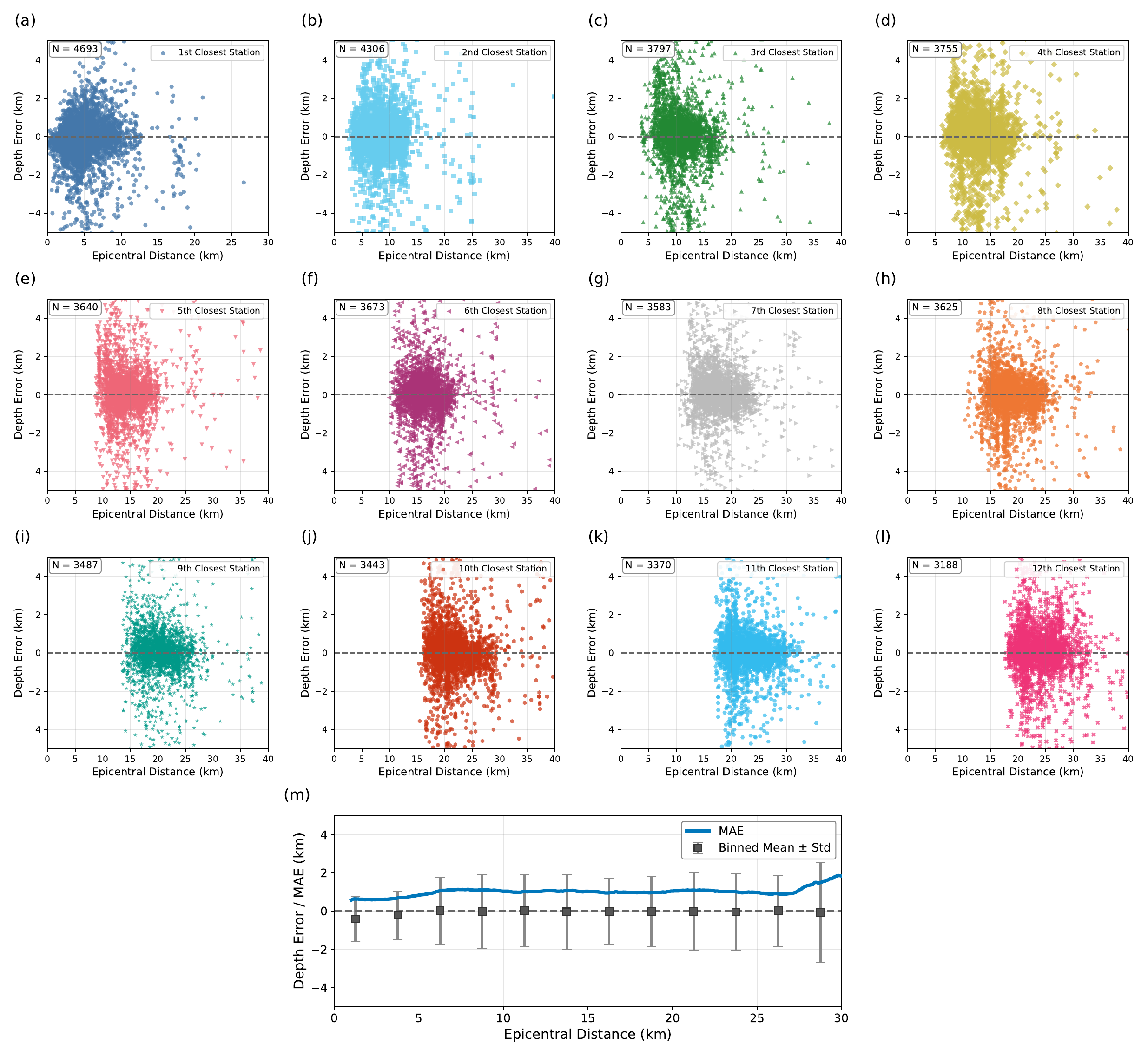}
\setlength{\abovecaptionskip}{3pt}  
\setlength{\belowcaptionskip}{1pt}

\caption{Analysis of depth errors with respect to the epicentral distances for the 12 closest stations (from 1st to 12th, panels a-l) per event. Panel (m) illustrates the mean absolute error (MAE) and binned mean standard deviation as a function of epicentral distances. (spaced every 2.5 km).}
\label{fig: closet_epi}
\end{figure}

With the regionally generalized model, depth performance can be further improved by averaging predictions across all available regional stations. To simulate real-world scenarios --- such as a lack of nearby stations --- we removed the nearest station (Figure~\ref{fig: ave_com_epi13}) and the two nearest stations (Figure~\ref{fig: ave_com_epi14}) and then re-estimated depths using the next 12 closest stations. By comparing these two scenarios with the original station setting (Figure~\ref{fig: ave_com_epi12}), we found that the influence on depth error and recall remained within 0.15 km and 5\%, respectively, indicating a negligible effect. The normalized recovery matrices (Figure~\ref{fig: ave_com_epi12}c,f,i,l; Figure~\ref{fig: ave_com_epi13}c,f,i,l; Figure~\ref{fig: ave_com_epi14}c,f,i,l) consistently show strong diagonal patterns across all three configurations, demonstrating robust depth resolution regardless of the nearest station availability. Excluding the two nearest stations --- thereby leaving most stations more than 5 km away --- further highlights that our deep learning method is weakly dependent on station proximity.

We repeated the same analysis for scenarios excluding the nearest station (Figure~\ref{fig: ave_com_epi13_AUCPR}) and the two nearest stations (Figure~\ref{fig: ave_com_epi14_AUCPR}). The results show comparable performance with AUC-PR values of 0.894 and 0.890, and F1-scores ranging from 0.857 to 0.882 (excluding one station) and 0.847 to 0.874 (excluding two stations). Station counts of 6-8 continue to achieve well-balanced F1-scores exceeding 0.865, confirming that the regionally generalized model maintains reliable depth prediction even when the closest stations are absent.

\begin{figure}[htpb]
\centering
\includegraphics[width=1\textwidth]{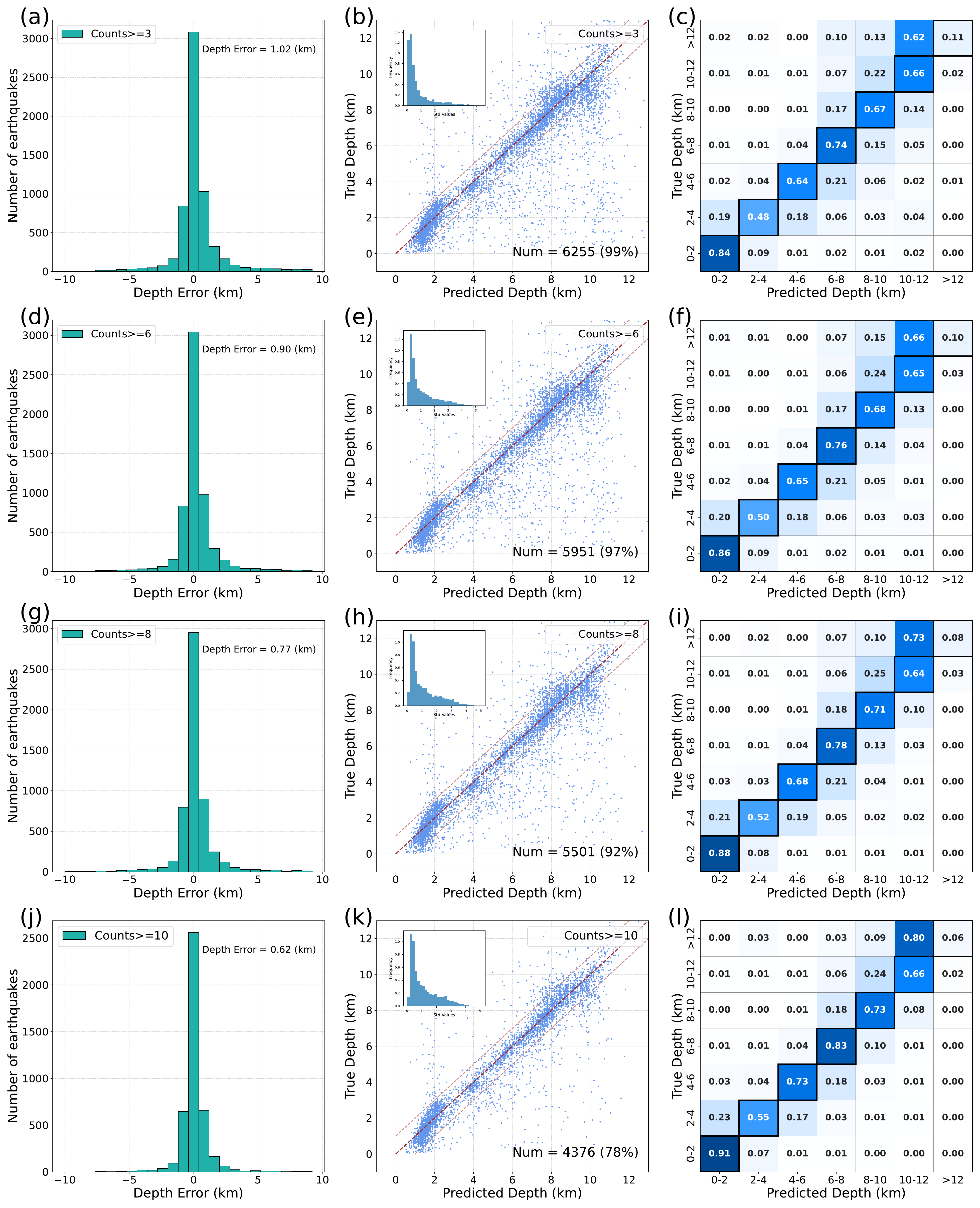}
\setlength{\abovecaptionskip}{3pt}  
\setlength{\belowcaptionskip}{1pt}
\caption{
Same as Figure~\ref{fig: ave_com_epi12}, except that depth prediction using the average of the 13 nearest stations, with the nearest station removed for each event. Thus, 12 stations still contribute to final depth determination. }
\label{fig: ave_com_epi13}
\end{figure}

\begin{figure}[htpb]
\centering
\includegraphics[width=1\textwidth]{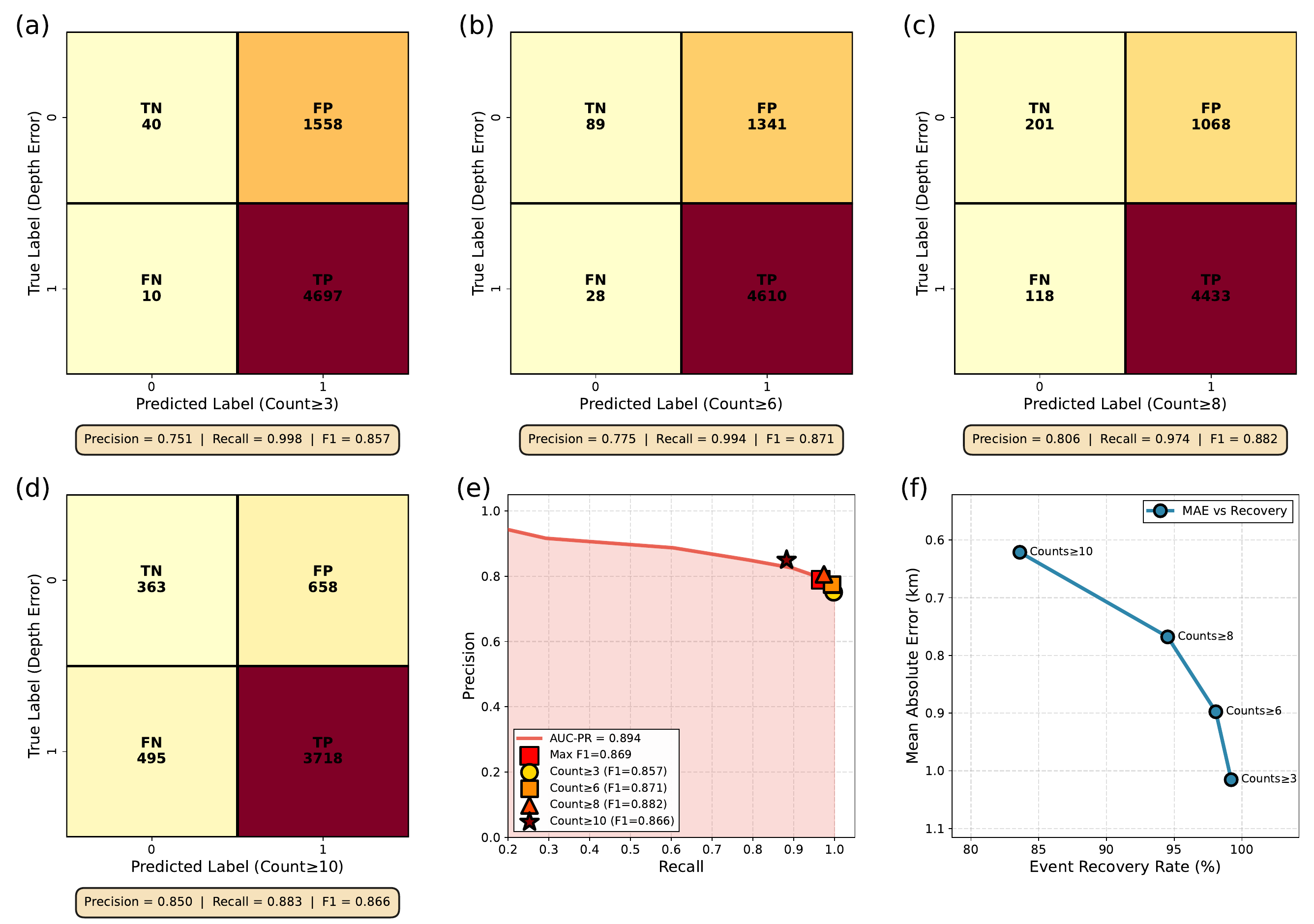}
\setlength{\abovecaptionskip}{3pt}  
\setlength{\belowcaptionskip}{1pt}
\caption{
Same as Figure~\ref{fig: ave_com_epi12_AUCPR}, except that depth prediction using the average of the 13 nearest stations, with the nearest station removed for each event. Thus, 12 stations still contribute to final depth determination. }
\label{fig: ave_com_epi13_AUCPR}
\end{figure}

\begin{figure}[htpb]
\centering
\includegraphics[width=1\textwidth]{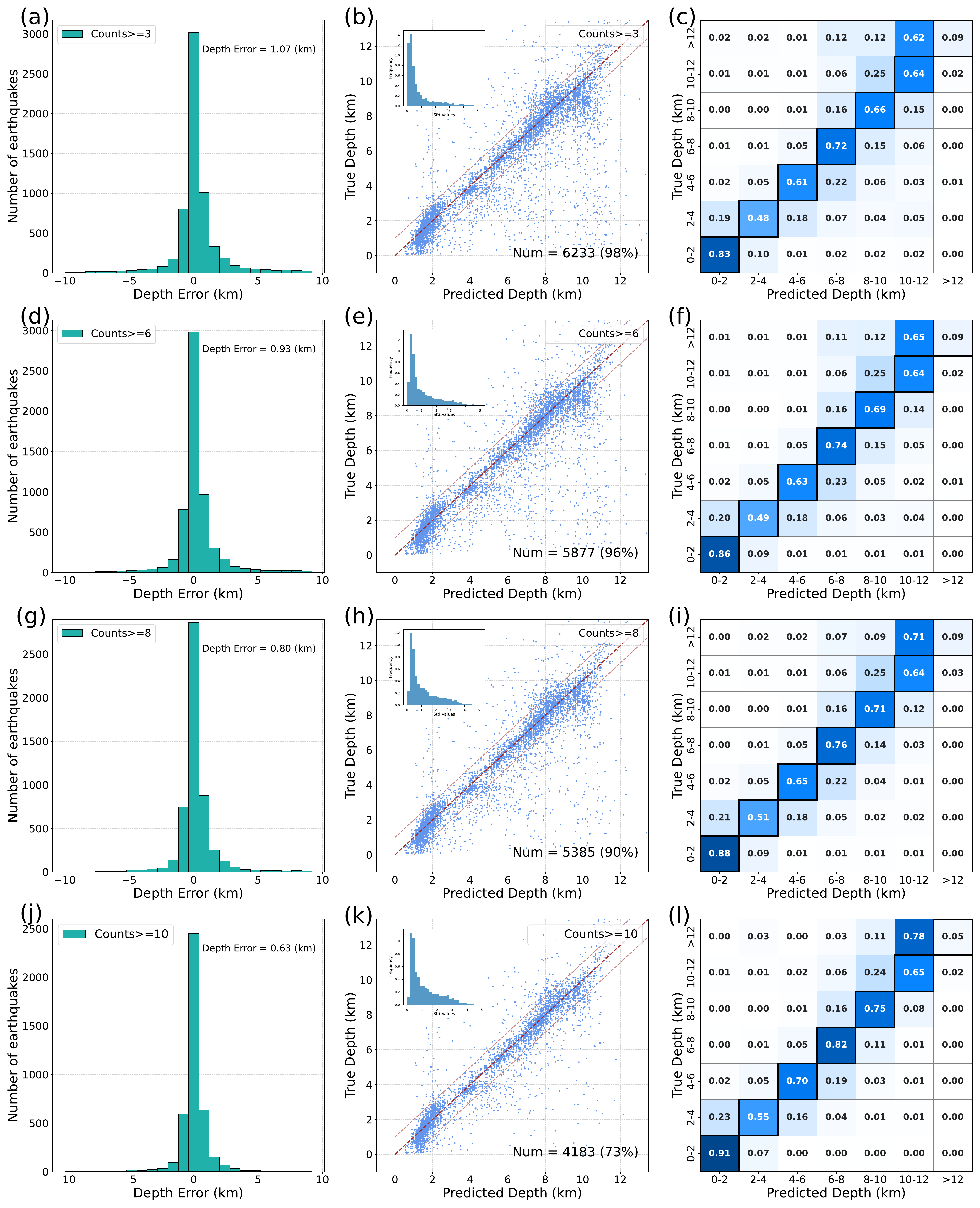}
\setlength{\abovecaptionskip}{3pt}  
\setlength{\belowcaptionskip}{1pt}
\caption{Same as Figure~\ref{fig: ave_com_epi12}, except that depth prediction using the average of the 14 nearest stations, with the nearest two stations removed for each event. Thus, 12 stations still contribute to final depth determination.  }
\label{fig: ave_com_epi14}
\end{figure}

\begin{figure}[htpb]
\centering
\includegraphics[width=1\textwidth]{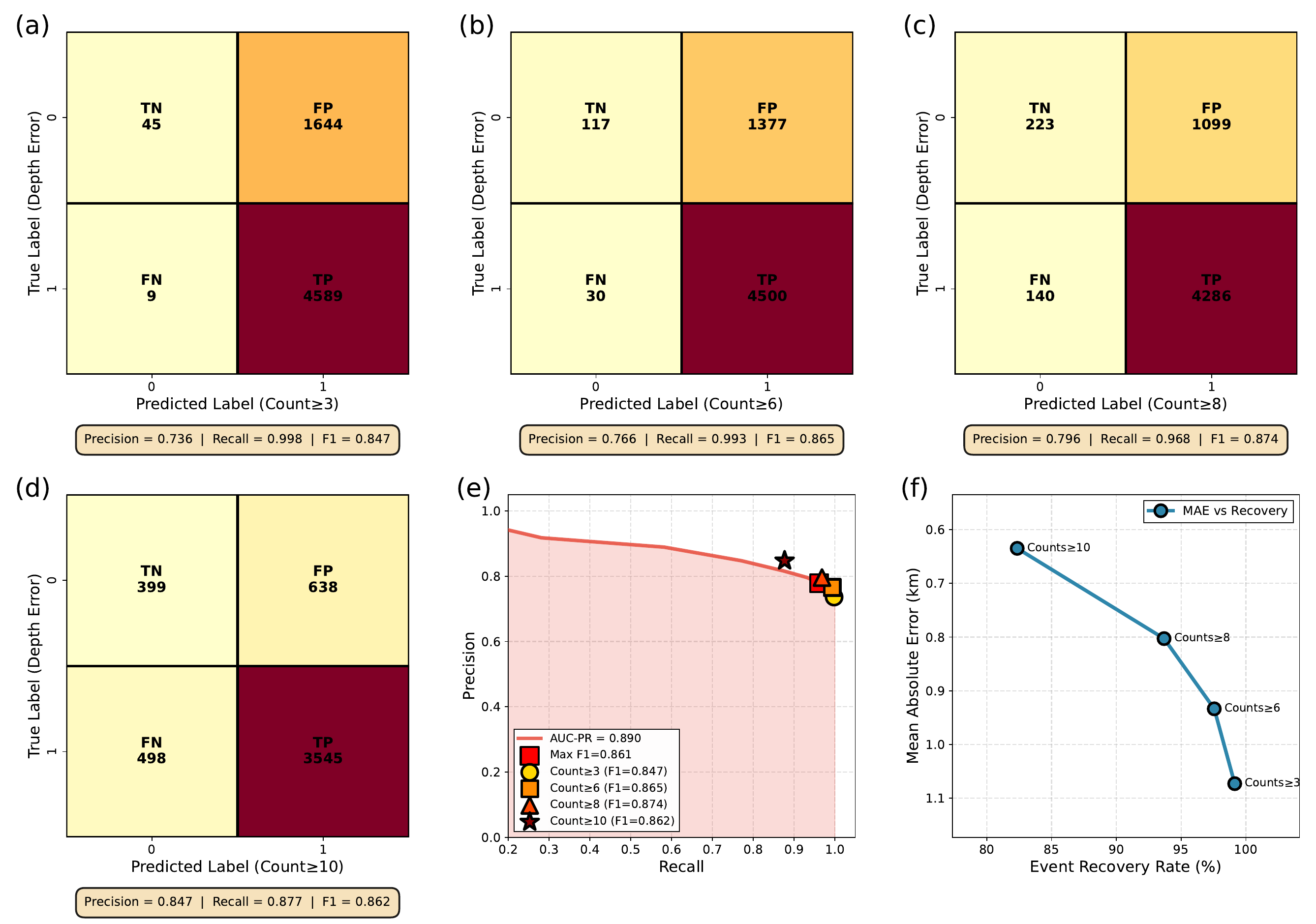}
\setlength{\abovecaptionskip}{3pt}  
\setlength{\belowcaptionskip}{1pt}
\caption{Same as Figure~\ref{fig: ave_com_epi12_AUCPR}, except that depth prediction using the average of the 14 nearest stations, with the nearest two stations removed for each event. Thus, 12 stations still contribute to final depth determination.  }
\label{fig: ave_com_epi14_AUCPR}
\end{figure}


\section{Discussion}

\subsection{  Depth Error Analysis as a Function of Magnitude and SNR}

In this subsection, we comprehensively evaluate our deep learning-based depth estimation network by quantifying its sensitivity to event magnitude and data quality (i.e., SNR). Here, a single station (i.e., CESI) model was used for analysis, with the threshold set to 0.95. As shown in Figure~\ref{fig: SNR_EPI}, larger-magnitude earthquakes and higher-SNR waveforms consistently yield lower depth errors --- an effect observed across all methods. Depth estimation error decreases with increasing magnitude  (as illustrated by the red dashed lines in Figure~\ref{fig: SNR_EPI}a), a behavior that should inform both data selection and model training. Depth accuracy is likewise strongly correlated with SNR: errors decline markedly as SNR increases (Figure~\ref{fig: SNR_EPI}b), highlighting the network's sensitivity to signal quality. High-SNR data, with distinct waveform signatures, facilitate robust extraction of depth information from continuous seismic waveforms, underscoring the critical role of denoising in preprocessing (e.g., filtering). Notably, when applying a prediction probability threshold of 0.98, depth errors no longer depend on magnitude or SNR (Figure~\ref{fig: SNR_EPI_0.98}a and b), demonstrating the model's reliability. However, raising the threshold lowers depth errors but also reduces event recovery rate. Thus, as with all techniques, high-quality waveform recordings --- obtained via nearby, densely spaced stations --- remain crucial for maximizing the model's performance.

\begin{figure}[htbp]
\centering
 \noindent\includegraphics[width=0.6\textwidth]{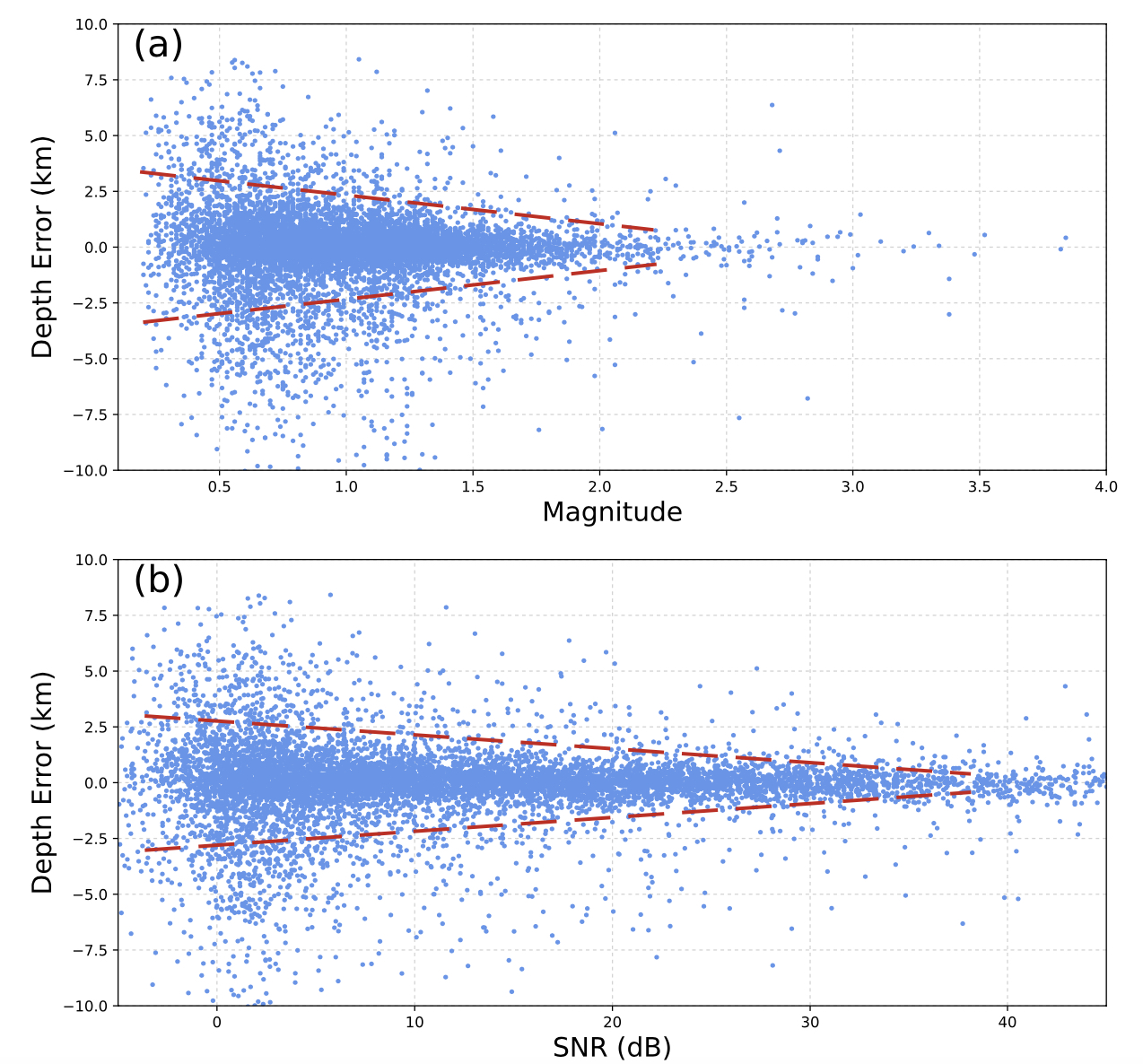} 
 \setlength{\abovecaptionskip}{2pt}  
\setlength{\belowcaptionskip}{5pt}
\caption {Depth error sensitivity analysis from individual stations based on the regionally generalized model with respect to different factors: magnitude, SNR (dB). Red dashed lines in panels a and b indicate that the depth errors decrease with increasing magnitude and SNR. 
}
\label{fig: SNR_EPI}
\end{figure}

\begin{figure}[htbp]
\centering
 \noindent\includegraphics[width=0.65\textwidth]{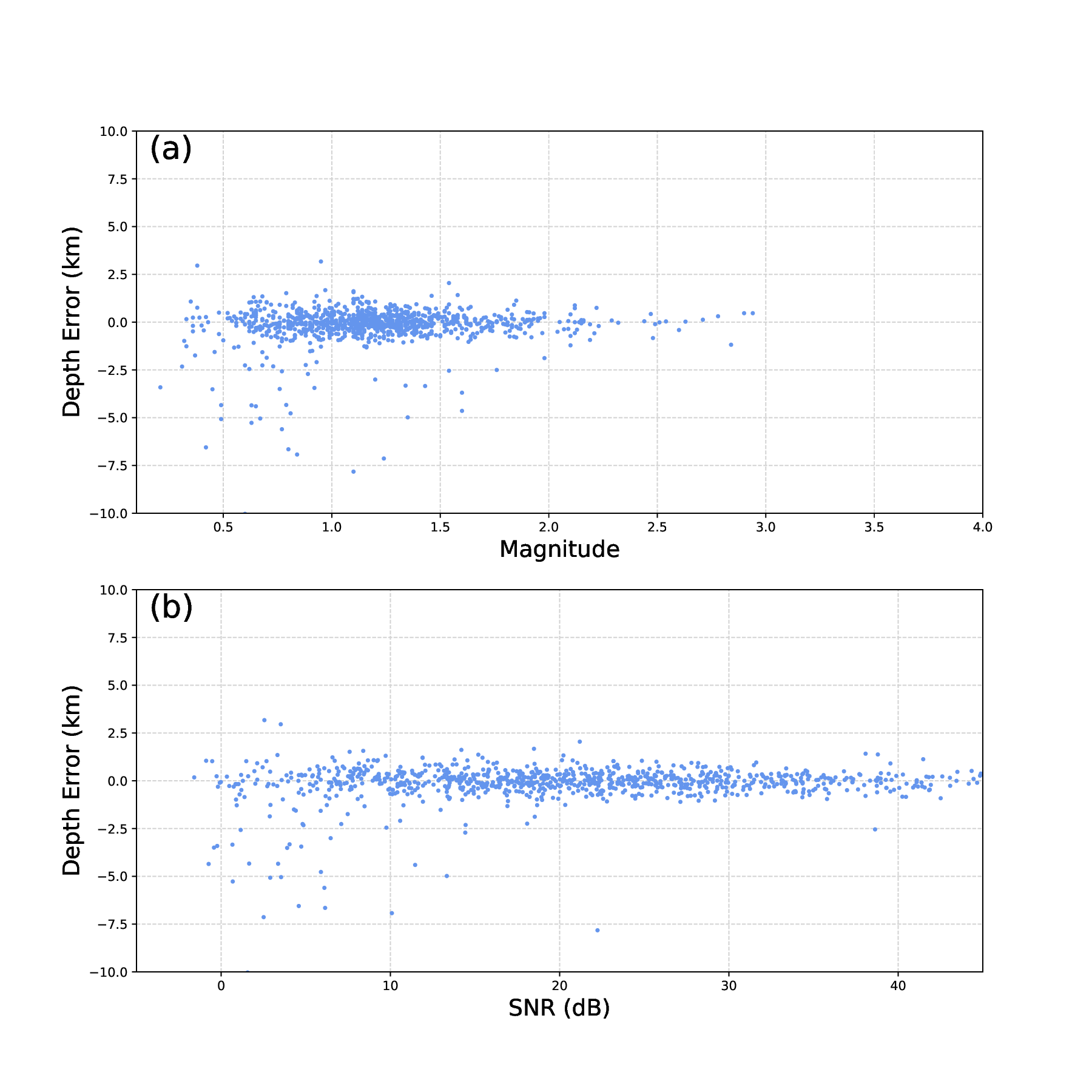} 
  \setlength{\abovecaptionskip}{2pt}  
\setlength{\belowcaptionskip}{5pt}
\caption {
Same as Figure~\ref{fig: SNR_EPI}, except that the threshold is 0.98 for the single station (CESI) network model's depth sensitivity analysis. }
\label{fig: SNR_EPI_0.98}
\end{figure}

\subsection{Relatively Large Depth Errors for Shallow Events ($<\ \sim$2 km)}

We observed that the neural network's depth predictions exhibit a noticeable upward distortion in the shallow depth range ($<\ \sim$2 km): depths at 0-1 km tend to be overestimated, while those around 2 km tend to be underestimated. This trend matches the field findings of \cite{zhang2022spatiotemporal}. However, their network --- trained on precisely labeled synthetic data --- does not show this bias, indicating that the errors arise from the labeling inaccuracies. Similarly, we speculate that inaccurate depth labels for shallow events in the field catalog --- perhaps due to topographic variations in this case --- are responsible for this discrepancy.

\subsection{ Transfer Learning for Pseudo-Newly Deployed Stations}

We evaluated the performance of a regionally generalized model for a pseudo-newly deployed station. We retrained the model using data from all stations except ED23 (Figure 2) and then applied it to predict depths at station ED23. The resulting depth error is 2.35 km, which is significantly higher than that of three nearby stations (Figure~\ref{fig: Generalized}). Unlike phase picking models  \cite[]{zhu2019phasenet, zhu2023ustc}, where the neural networks are sensitive only to first arrivals, our depth neural network utilizes the entire waveform and learns complex structure information that constrains depth, leading to reduced regional transferability. However, transfer learning can substantially improve performance with only a small amount of labeled data \cite[]{chai2020using, niksejel2024obstransformer}. By transfer learning based on 5,000 and 10,000 samples at station ED23, we achieved acceptable errors of 1.38 km and 1.22 km, respectively (Figure S3a and c). Optimal performance stabilizes at 20,000 samples, yielding a depth error of 1.02 km (Figure S3e), comparable to the three nearby stations (Figure~\ref{fig: Generalized}). These demonstrate that the regionally generalized model enables us to obtain preliminary results for newly deployed stations  --- with a compromised accuracy  ---  yet transfer learning with a modest sample size markedly enhances depth predictions.

\subsection{ Applications to Sparse Networks and Historical Earthquakes}

While our models were trained using a relatively dense network, this approach enables an important application: transferring depth constraints from modern, well-instrumented periods to historical, sparse-network periods. This makes the method particularly suitable for refining depths of historical earthquakes recorded with single or a few stations (e.g., the pseudo-sparse station scenario: single-station prediction and three-station averages).

\subsection{ Limitations and Future Work}

We acknowledge that focal mechanism variability can affect waveform characteristics, which may in turn introduce uncertainties to depth estimation. However, our method utilizes the entire waveform, including scattered waves, which may not be significantly dependent on the focal mechanism. Such refined tests quantifying the relationship between focal mechanisms and depth estimation accuracy could be explored in future research.

Amplitude normalization was adopted during model training and testing, potentially eliminating magnitude differences. This elimination works well for small events (e.g., $M < 3$; Figure~\ref{fig: SNR_EPI},\ref{fig: SNR_EPI_0.98}), which can be approximated as point-sources. However, we acknowledge that depth-prediction accuracy for larger events may decrease due to rupture dimensions and the limited number of large events in our training dataset, and this analysis cannot be thoroughly investigated because large events are scarce in the field.

\section{Conclusion}

In this study, we developed a novel deep learning framework, VGGDepth, for accurately determining earthquake source depths directly from seismic waveforms. By adapting the VGG16 architecture from computer vision to process single-station three-component waveform data, our method provides a robust and efficient mechanism for mapping seismic waveforms to source depth estimates. We validated our method using real-world seismic data from the 2016-2017 Central Apennines, Italy earthquake sequence. Two models were introduced: one trained on individual single stations, and a second, regionally generalized model capable of handling waveforms from different stations within the target area. Both models achieved earthquake depth prediction accuracies of less than 1 km, with event recovery rates exceeding 71\%-85\% and F1-scores above 0.8. Comprehensive evaluation using confusion matrices, precision-recall analysis, and normalized recovery matrices demonstrated robust performance with AUC-PR values exceeding 0.80. To fully utilize available stations in the region, we also presented a strategy to improve the depth prediction by integrating multiple stations using the generalized model. Multi-station averaging further reduced depth errors to 0.6-0.8 km while maintaining high recovery rates. We further demonstrated that the generalized model has great potential to be applicable to newly deployed stations, and transfer learning can be used to improve their performance in such cases.

VGGDepth addresses long-standing challenges in depth determination based on first arrivals --- specifically, the poor sensitivity of first arrivals to depth, the trade-off between earthquake depth and origin time, and the strong dependence on the nearest stations --- particularly in regions with sparse or uneven station coverage. Moreover, compared to conventional methods based on depth-phase identification, VGGDepth offers advantages in efficiency and convenience by eliminating the need for intermediate calculations. This method holds great potential for refining the depths of historical earthquakes, especially during periods of limited station coverage, and for estimating source depths of seismic events on other planets, such as the Moon or Mars.

\section*{Acknowledgments}
We are grateful to the Editor Yangkang Chen, Honn Kao, Zhe Jia, and an anonymous reviewer for their insightful review and constructive suggestions. The authors thank Gregory Beroza and Zhigang Peng for valuable discussions.  This research was supported by the Natural Sciences and Engineering Research Council of Canada Discovery Grant (Grant Number RGPIN-2019-04297).

\section*{Conflict of Interest Statement}
The authors have no conflicts of interest to disclose.

\section*{Open Research}

Waveform data were downloaded from the Earthscope Inc. Data Management Center (DMC), formerly known as Incorporated Research Institutions for Seismology (IRIS) DMC, and from the National Institute of Geophysics and Volcanology (INGV) Data Centers. The earthquake catalog used in this study is from \citet{tan2021machine}.
 Figures in this article were generated using the Generic Mapping Tools \cite[]{wessel2013generic} and Matplotlib (visualization with Python, \citeauthor{Hunter:2007}, \citeyear{Hunter:2007}). The VGGDepth package has been released on Zenodo \cite[]{wenda2025code} 

%
%

\end{justify}

\bibliography{example.bib}

%
%
%
%
%

\end{document}